\documentclass[aps,superscriptaddress,11pt,twoside]{revtex4}

\usepackage{amsmath,latexsym,amssymb,verbatim, enumerate, graphicx}

\usepackage{theorem}
\newtheorem{definition}{Definition}[section]

\newtheorem{lemma}[definition]{Lemma}

\newtheorem{theorem}[definition]{Theorem}

\def\squareforqed{\hbox{\rlap{$\sqcap$}$\sqcup$}}
\def\qed{\ifmmode\squareforqed\else{\unskip\nobreak\hfil
\penalty50\hskip1em\null\nobreak\hfil\squareforqed
\parfillskip=0pt\finalhyphendemerits=0\endgraf}\fi}
\def\endenv{\ifmmode\;\else{\unskip\nobreak\hfil
\penalty50\hskip1em\null\nobreak\hfil\;
\parfillskip=0pt\finalhyphendemerits=0\endgraf}\fi}
\newenvironment{proof}{\noindent \textbf{{Proof~} }}{\qed}

\font\gensymbols=drgen10
\def\male{{\gensymbols\char"1A}}
\def\female{{\gensymbols\char"19}}

\mathchardef\ordinarycolon\mathcode`\:
\mathcode`\:=\string"8000
\def\vcentcolon{\mathrel{\mathop\ordinarycolon}}
\begingroup \catcode`\:=\active
  \lowercase{\endgroup
  \let :\vcentcolon
  }

\newcommand{\nc}{\newcommand}
\nc{\rnc}{\renewcommand} \nc{\beq}{\begin{equation}}
\nc{\eeq}{{\end{equation}}} \nc{\bea}{\begin{eqnarray}}
\nc{\eea}{\end{eqnarray}} \nc{\beqa}{\begin{eqnarray}}
\nc{\eeqa}{\end{eqnarray}} \nc{\lbar}[1]{\overline{#1}}
\nc{\bra}[1]{\langle#1|} \nc{\ket}[1]{|#1\rangle}
\nc{\ketbra}[2]{|#1\rangle\!\langle#2|}
\nc{\braket}[2]{\langle#1|#2\rangle} \nc{\proj}[1]{|
#1\rangle\!\langle #1 |} \nc{\avg}[1]{\langle#1\rangle}
\rnc{\max}{\operatorname{max}} \nc{\rank}{\operatorname{rank}}
\nc{\conv}{\operatorname{conv}}
\nc{\smfrac}[2]{\mbox{$\frac{#1}{#2}$}}
\nc{\Tr}{\operatorname{Tr}}
\nc{\ox}{\otimes}
\nc{\dg}{\dagger}
\nc{\dn}{\downarrow} \nc{\cA}{{\cal A}} \nc{\cB}{{\cal B}}
\nc{\cC}{{\cal C}} \nc{\cD}{{\cal D}} \nc{\cE}{{\cal E}}
\nc{\cF}{{\cal F}} \nc{\cG}{{\cal G}} \nc{\cH}{{\cal H}}
\nc{\cI}{{\cal I}} \nc{\cJ}{{\cal J}} \nc{\cK}{{\cal K}}
\nc{\cL}{{\cal L}} \nc{\cM}{{\cal M}} \nc{\cN}{{\cal N}}
\nc{\cO}{{\cal O}} \nc{\cP}{{\cal P}} \nc{\cR}{{\cal R}}
\nc{\cS}{{\cal S}} \nc{\cT}{{\cal T}} \nc{\cU}{{\cal U}}
\nc{\cX}{{\cal X}} \nc{\cW}{{\cal W}} \nc{\cZ}{{\cal Z}}
\nc{\csupp}{{\operatorname{csupp}}}
\nc{\qsupp}{{\operatorname{qsupp}}} \nc{\var}{\operatorname{var}}
\nc{\rar}{\rightarrow} \nc{\lrar}{\longrightarrow}
\nc{\poly}{\operatorname{poly}}
\nc{\polylog}{\operatorname{polylog}}
\nc{\Lip}{\operatorname{Lip}} \nc{\1}{\openone}

\def\>{\rangle}
\def\<{\langle}

\def\a{\alpha}

\def\d{\delta}
\def\e{\epsilon}

\def\r{\rho}
\def\s{\sigma}

\def\ph{\varphi}

\def\ps{\psi}

\def\Ph{\Phi}

\nc{\glneq}{{\raisebox{0.6ex}{$>$}  \hspace*{-1.8ex} \raisebox{-0.6ex}{$<$}}}
\nc{\gleq}{{\raisebox{0.6ex}{$\geq$}\hspace*{-1.8ex} \raisebox{-0.6ex}{$\leq$}}}

\nc{\RR}{{{\mathbb R}}}
\nc{\CC}{{{\mathbb C}}}
\nc{\FF}{{{\mathbb F}}}
\nc{\HH}{{{\mathbb H}}}
\nc{\NN}{{{\mathbb N}}}
\nc{\ZZ}{{{\mathbb Z}}}
\nc{\PP}{{{\mathbb P}}}
\nc{\QQ}{{{\mathbb Q}}}
\nc{\UU}{{{\mathbb U}}}
\nc{\WW}{{{\mathbb W}}}
\nc{\EE}{{{\mathbb E}}}
\rnc{\SS}{{{\mathbb S}}}
\nc{\id}{{\operatorname{id}}}

\nc{\vholder}[1]{\rule{0pt}{#1}}

\nc{\ob}[1]{#1}

\def\beq{\begin {equation}}
\def\eeq{\end {equation}}

\nc{\eq}[1]{Eq.~(\ref{eq:#1})} \nc{\eqs}[2]{Eqs.~(\ref{eq:#1}) and
(\ref{eq:#2})}

\nc{\eqn}[1]{Eq.~(\ref{eqn:#1})}
\nc{\eqns}[2]{Eqs.~(\ref{eqn:#1}) and (\ref{eqn:#2})}

\nc{\region}{\cS\cW}

\begin{document}

\title{{\Large The mother of all protocols:\protect\\
               Restructuring quantum information's family tree}}

\author{Anura Abeyesinghe}
 \email{anura@caltech.edu}
 \affiliation{
    Institute for Quantum Information,
    Physics Department, Caltech 103-33, Pasadena, CA
    91125, USA
    }
\author{Igor Devetak}
 \email{devetak@usc.edu}
 \affiliation{
    Electrical Engineering Department,
    University of Southern California
    Los Angeles, California 90089, USA
    }
\author{Patrick Hayden}
 \email{patrick@cs.mcgill.ca}
 \affiliation{
    School of Computer Science,
    McGill University,
    Montreal, Quebec, H3A 2A7, Canada
    }
\author{Andreas Winter}
 \email{a.j.winter@bris.ac.uk}
 \affiliation{
    Department of Mathematics,
    University of Bristol,
    University Walk, Bristol BS8 1TW, U.~K.
    }

\date{June 27, 2006}

\begin{abstract}
We give a simple, direct proof of the ``mother'' protocol of
quantum information theory. In this new formulation, it is easy to
see that the mother, or rather her generalization to the fully
quantum Slepian-Wolf protocol, simultaneously accomplishes two
goals: quantum communication-assisted entanglement distillation,
and state transfer from the sender to the receiver. As a result,
in addition to her other ``children,'' the mother protocol
generates the state merging primitive of Horodecki, Oppenheim and
Winter, a fully quantum reverse Shannon theorem, and a new class
of distributed compression protocols for correlated quantum
sources which are optimal for sources described by separable
density operators. Moreover, the mother protocol described here is
easily transformed into the so-called ``father'' protocol whose
children provide the quantum capacity and the
entanglement-assisted capacity of a quantum channel, demonstrating
that the division of single-sender/single-receiver protocols into
two families was unnecessary: all protocols in the family are
children of the mother.
\end{abstract}

\maketitle

\parskip .75ex


\section{Introduction} \label{sec:intro}

One of the major goals of quantum information theory is to find
the optimal ways to make use of noisy quantum states or channels
for communication or establishing entanglement. Quantum Shannon
theory attacks the problem in the limit of many copies of the
state or channel in question, in which situation the answers often
simplify to the point where they can be expressed by relatively
compact formulae. The last ten years have seen major advances in
the area, including, among many other discoveries, the
determination of the classical capacity of a quantum
channel~\cite{H98,SW97}, the capacities of entanglement-assisted
channels~\cite{BSST99,BSST02}, the quantum capacity of a quantum
channel~\cite{L96,S02,D05}, and the best ways to use noisy
entanglement to extract pure entanglement~\cite{DW05} or to help
send classical information~\cite{HHHLT01}. Until recently,
however, each new problem was solved essentially from scratch and
no higher-level structure was known connecting the different
results. Harrow's introduction of the \emph{cobit}~\cite{H04} and
its subsequent application to the construction of the so-called
``mother'' and ``father'' protocols provided that missing
structure. Almost all the problems listed above were shown to fall
into two families, first the mother and her descendants, and
second the father and his~\cite{DHW04}. Appending or prepending
simple transformations like teleportation and superdense coding
sufficed to transform the parents into their children.

In this paper, we provide a direct proof of the mother protocol
or, more precisely, of the existence of a protocol performing the
same task as the mother. In contrast to most proofs in information
theory, instead of showing how to establish perfect correlation of
some kind between the sender (Alice) and the receiver (Bob), our
proof proceeds by showing that the protocol \emph{destroys} all
correlation between the sender and a reference system. Since
destruction is a relatively indiscriminate goal, the resulting
proof is correspondingly simple. This approach also makes it clear
that the mother actually accomplishes more than originally
thought. In particular, in addition to distilling entanglement
between Alice and Bob, the protocol transfers all of Alice's
entanglement with a reference system to Bob. This side effect is
very important in its own right, and a major focus of our paper.
To start, it places the state merging protocol of Horodecki,
Oppenheim and Winter~\cite{HOW05,HOW05b} squarely within the
mother's brood. In addition, it makes it possible to use the
mother as a building block for distributed compression. We analyze
the resulting protocols, finding they are optimal for sources
described by separable density operators, as well as inner and
outer bounds on the achievable rate region in general.

We also emphasize a further connection, first identified
in~\cite{D05b}, that requires both the state transfer and
entanglement distillation capabilities of the mother: the entire
protocol allows for a time-reversed interpretation as a quantum
reverse Shannon theorem, that is, an efficient simulation of a
noisy quantum channel using a noiseless quantum channel along with
entanglement.

Finally, the new approach to the mother solves a major problem
left unanswered in the original family paper. There, no
operational relationship between the mother and father protocols
could be identified, but the two were nonetheless connected by a
formal symmetry called \emph{source-channel duality}~\cite{D05b}.
This new mother protocol can be directly transformed into the
father, resolving the mystery of the two parents' formal
similarity and collapsing the two families into one.

The structure of the paper is as follows. After reviewing the
family of quantum protocols in section~\ref{sec:family} and
providing, in section~\ref{sec:protocol} a high-level description
of the improved mother, henceforth the fully quantum Slepian-Wolf
(FQSW) protocol, we go straight to the statement and proof of the
central result of this paper in section~\ref{sec:oneShotMother}: a
one-shot version of FQSW. The middle section of the paper is
devoted to a number of applications of one-shot FQSW.
Sections~\ref{sec:father_1} and~\ref{sec:qrst-oneshot} describe
one-shot versions of the ``father'' and the fully quantum
``quantum reverse Shannon'' (FQRS) protocol, respectively. The
one-shot theorems quickly yield memoryless forms for all three:
FQSW in section~\ref{sec:iid}, the father in
section~\ref{sec:father:iid} and FQRS in
section~\ref{sec:qrst-iid}. Then we turn to the other highlight of
this paper, a treatment of the fully quantum version of the
distributed compression problem, which we can solve completely for
a large class of sources by providing general inner and outer
bounds on the rate region, in section~\ref{sec:distributed}. In
section~\ref{sec:efficiency} we point out that the FQSW protocol
allows for efficient encoding via Clifford operations, after which
we conclude. An appendix collects useful facts on typical
subspaces.

{\bf Notation:} For a quantum system $A$, let $d_A = \dim A$. For
two quantum systems $A$ and $A'$, let $F^{A}$ be the operator that
swaps the two systems. An operator acting on a subsystem is freely
identified with its extension (via tensor product with the
identity) to larger systems. $\Pi^{A}_{+}$ denotes the projector
onto the symmetric subspace of $A \ox A'$ and $\Pi^A_{-}$ the
projector onto the antisymmetric subspace of $A \ox A'$. Let
$\UU(A)$ be the unitary group on $A$. $H(A)_\ph$ is the von
Neumann entropy of $\ph^A$, $I(A;B)_\ph = H(A)_\ph + H(B)_\ph -
H(AB)_\ph$ is the mutual information between the $A$ and $B$ parts
of $\ph$ and $H(A|B)_\ph = H(AB)_\ph - H(B)_\ph$ the conditional
entropy. The symbol $\ket{\Ph}^{AB}$ will be used to represent a
maximally entangled state between $A$ and $B$. Logarithms are
taken base 2 throughout.

\section{The family of quantum protocols} \label{sec:family}

The mother protocol is a transformation of a tensor power quantum
state $(\ket{\ph}^{ABR})^{\ox n}$. At the start, Alice holds the
$A$ shares and Bob the $B$ shares. $R$ is a reference system
purifying the $AB$ systems and does not participate actively in
the protocol. In the original formulation, the mother protocol
accomplished a type of entanglement distillation between Alice and
Bob in which the only communication permitted was the ability to
send \emph{qubits} from Alice to Bob. The transformation can be
expressed concisely in the resource inequality formalism as
\begin{equation} \label{eqn:mother}
 \langle \ph^{AB} \rangle + \frac{1}{2} I(A;R)_\ph \, [q\rar q]
 \geq
 \frac{1}{2} I(A;B)_\ph \, [qq].
\end{equation}
We will informally explain the resource inequalities used here,
but the reader is directed to~\cite{DHW05} for a rigorous
treatment. $[q \rar q]$ represents one qubit of communication from
Alice to Bob and $[qq]$ represents an ebit shared between them. In
words, $n$ copies of the state $\ph$ shared between Alice and Bob
can be converted into $I(A;B)_\ph$ EPR pairs per copy provided
Alice is allowed to communicate with Bob by sending him qubits at
rate $I(A;R)_\ph$ per copy. Small imperfections in the final state
are permitted provided they vanish as $n$ goes to infinity.

\begin{figure}
 \begin{center}
  \includegraphics[width=5in,bbllx=8,bblly=430,bburx=784,bbury=620]{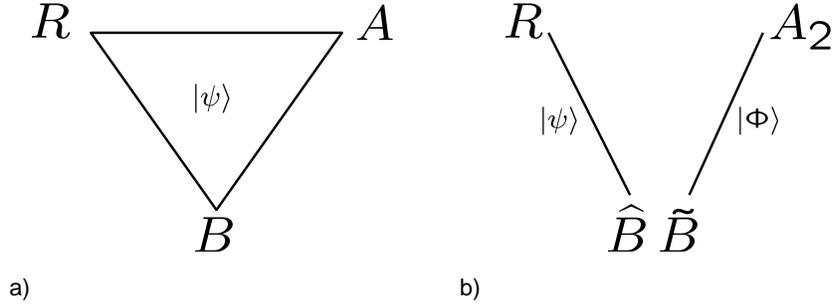}
 \end{center}
 \caption{a) The starting point for the FQSW protocol, a pure
 tripartite entangled state $\ket{\ps} = (\ket{\ph}^{ABR})^{\ox n}$.
 b) After execution of the protocol, Alice's portion of the original
 tripartite state has been transferred to Bob, so that Bob holds a purification
 of the unchanged reference system in his register $\widehat{B}$. He also
 shares pure state entanglement with Alice in the form of the state $\ket{\Phi}$.}
 \label{fig:FQSW}
\end{figure}

In this paper, we prove a stronger resource inequality which we
call the \emph{fully quantum Slepian-Wolf} (FQSW) inequality. The
justification for this name will become apparent in Section
\ref{sec:distributed}, where we study its applicability to
distributed compression, solved classically by Slepian and Wolf
\cite{SW71}. The inequality states that starting from state
$(\ket{\ph}^{ABR})^{\ox n}$ and using only quantum communication
at the rate $\smfrac{1}{2} I(A;R)_\ph$ from Alice to Bob, they can
distill EPR pairs at the rate $\smfrac{1}{2} I(A;B)_\ph$ and
produce a state approximating $(\ket{\ps}^{R\widehat{B}})^{\ox
n}$, where $\widehat{B}$ is held by Bob and $\ph^R = \ps^R$. That
is, Alice can \emph{transfer} her entanglement with the reference
system $R$ to Bob while simultaneously distilling ebits with him.
A graphical depiction of this transformation is given in Figure
\ref{fig:FQSW}. The process can also be expressed as a resource
inequality in the following way:
\begin{equation} \label{eqn:FQSW}
 \langle W^{S \rar AB} : \ph^S \rangle
 + \frac{1}{2} I(A;R)_\ph \, [q \rar q]
 \geq
 \frac{1}{2} I(A;B)_\ph \, [qq]
 + \langle \id^{S \rar \widehat{B}} : \ph^S \rangle
\end{equation}
This inequality makes use of the concept of a relative resource. A
resource of the form $\langle \mathcal{N} : \r^S \rangle$ is a
channel with input system $S$ that is guaranteed to behave like
the channel $\mathcal{N}$ provided the reduced density operator of
the input state on $S$ is $\r^S$. In the inequality,
$W^{S \rar AB}$ is an isometry taking the $S$ system to
$AB$. Thus, on the left hand side of the inequality, a state is
distributed to Alice and Bob while on the right hand side, that
same state is given to Bob alone. Transforming the first situation
into the second means that Alice transfers her portion of the
state to Bob.

Since the relationship of the mother to entanglement distillation
and communication supplemented using noisy entanglement is
explained at length in the original family paper, we will not
describe the connections here. The FQSW inequality is stronger
than the mother, however, and leads to more children. In
particular, if the entanglement produced at the end of the
protocol is then re-used to perform teleportation, we get the
following resource inequality:
\begin{equation}
 \langle W^{S \rar AB} : \ph^S \rangle
 + H(A|B)_\ph\,[q\rar q] + I(A;B)_\ph \, [c\rar c]
 \geq
 \langle \id^{S \rar \widehat{B}} : \ph^S \rangle,
 \label{eqn:merging}
\end{equation}
which is known as the \emph{state merging} primitive~\cite{HOW05}.
It is of note both because it is a useful building block for
multiparty protocols~\cite{HOW05,HOW05b,YHD06} and because it
provides an operational interpretation of the conditional entropy
$H(A|B)_\ph$ as the number of qubits Alice must send Bob in order
to transfer her state to him, ignoring the classical communication
cost.

On the other side of the family there is the father protocol. In
contrast to the mother, in which Alice and Bob share a mixed state
$(\ph^{AB})^{\ox n}$, for the father protocol they are connected by a noisy
channel $\cN^{A'\rar B}$. Let $U^{A' \rar BE}$ be a Stinespring
dilation of $\cN$ with environment system $E$, such that
$\cN(\rho) = \Tr_E U \rho U^\dg$, and define $\ket{\ph}^{ABE} =
U^{A' \rar BE}\ket{\ph}^{AA'}$ for a pure state $\ket{\ph}^{AA'}$.
The resource inequality is
\begin{equation} \label{eqn:father}
 \langle \cN^{A' \rar B} \rangle
 + \frac{1}{2} I(A;E)_\ph \, [qq]
 \geq
 \frac{1}{2} I(A;B)_\ph \, [q \rar q].
\end{equation}
Thus, Alice and Bob use pre-existing shared entanglement and the
noisy channel to produce noiseless quantum communication.
Comparing Eq.~(\ref{eqn:father}) to the mother,
Eq.~(\ref{eqn:mother}), reveals the two to be strikingly similar:
to go from one to the other it suffices to replace channels by
states and vice-versa, as well as replace the reference $R$ by the
environment $E$. This formal symmetry is known as source-channel
duality~\cite{D05b}. Just as the mother can be strengthened to the
fully quantum Slepian-Wolf protocol, there is a fully coherent
version of the father known as the \emph{feedback
father}~\cite{D05b}.

\begin{figure}
 \begin{center}
  \includegraphics[width=5in,bbllx=8,bblly=450,bburx=784,bbury=700]{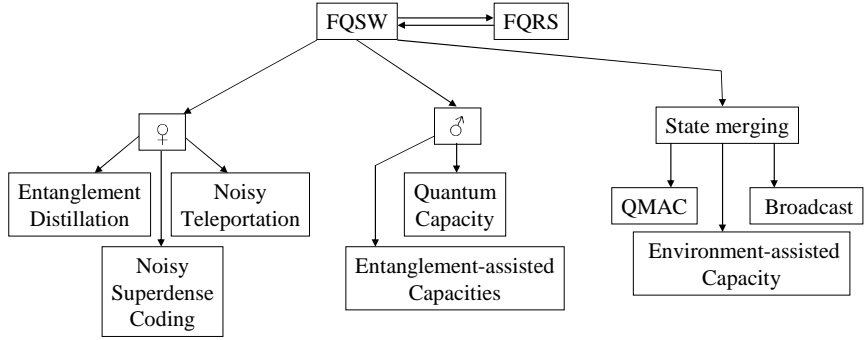}
 \end{center}
 \caption{Partial quantum information theory family tree. The symbols \female\, and
 \male\,
 represent the ``old'' mother and father protocols from \cite{DHW04} and arrows indicate
 that a protocol accomplishing the task at the start of the arrow can be transformed into
 a protocol accomplishing the task at the end. The relationships between \female\, , \male\, and
 their children are discussed in detail in \cite{DHW04,DHW05}. ``QMAC'' refers to the task of sending quantum data
 through a quantum multiple access channel~\cite{HOW05,YDH05}, ``broadcast'' the task of sending
 quantum data through a quantum broadcast channel~\cite{YHD06} and
 the environment-assisted quantum capacity is discussed
 in~\cite{SVW05}.}
 \label{fig:tree}
\end{figure}

The relationships between different protocols are sketched as a
family tree in Figure \ref{fig:tree}.

\section{The fully quantum Slepian-Wolf protocol} \label{sec:protocol}

The input to the fully quantum Slepian-Wolf protocol is a quantum
state, $(\ket{\ph}^{RAB})^{\ox n}$, and the output is also a
quantum state, $\ket{\Ph}^{A_2 \tilde{B}}
(\ket{\ph}^{R\widehat{B}})^{\ox n}$. $A_2$ is a quantum system
held by Alice while both $\tilde{B}$ and $\widehat{B}$ are held by
Bob. $\ket{\Ph}^{A_2 \tilde{B}}$ therefore represents a maximally
entangled state shared between Alice and Bob. The size of the
$A_2$ system is $nI(A;B)_\ph - o(n)$ qubits. The steps in the
protocol that transform the input state to the output state are as
follows:
\begin{enumerate}
\item Alice performs Schumacher compression on her system $A^n$.
The output space $A_S$ factors into two subsystems $A_1$ and $A_2$
with $\log d_{A_1} = n I(A;R) + o(n)$.
\item Alice applies a unitary transformation $U_A$ to $A_S$ and
then sends $A_1$ to Bob.
\item Bob applies an isometry $V_B$ taking $A_1 B^n$ to
$\widehat{B}\tilde{B}$.
\end{enumerate}
It remains to specify which transformations $U_A$ and $V_B$ Alice
and Bob should apply, as well as a more precise bound on
$d_{A_1}$. Observe that each step in the protocol is essentially
non-dissipative. Since essentially no information is leaked to the
environment at any step, Bob will hold a purification of the $A_2
R^n$ system after step $2$, regardless of the choice of $U_A$.
Because all purifications are equivalent up to local isometric
transformations of the purifying space, it therefore suffices to
ensure that the reduced state on $A_2 R^n$ approximates $\Ph^{A_2}
\ox (\ph^{R})^{\ox n}$ after step $2$. Bob's isometry $V_B$ will
be the one taking the purification he holds upon receiving $A_2$
to the one approximating $\ket{\Ph}^{A_2 \tilde{B}}
(\ket{\ph}^{R\widehat{B}})^{\ox n}$.

From this perspective, the operation $\r \rar \Tr_{A_1} ( U_A \,
\r \, U_A^\dg )$ should be designed to \emph{destroy} the
correlation between $A_2$ and $R^n$: the mother will succeed
provided the state on $A_2 \ox R^n$ is a product state and $A_2$
is maximally mixed. The operation $U_A$ does not itself destroy
the correlation; the partial trace over $A_1$ does that. $U_A$
should therefore be chosen in order to ensure that tracing over
$A_1$ should be maximally effective. Because one qubit can carry
at most two bits of information, tracing over a qubit can reduce
mutual information by at most two bits. The starting state
$(\ph^{AR})^{\ox n}$ has $nI(A;R)_\ph$ bits of mutual information,
which means that $A_1$ must consist of at least $\smfrac{n}{2}
I(A;R)_\ph$ qubits. We will see that by choosing $U_A$ randomly
according to the Haar measure we will come close to achieving this
rate.

The result is similar in spirit to a recent result of Groisman et
al. that demonstrated that in order to destroy correlation in the
state $\ph$ by discarding \emph{classical} information instead of
quantum, Alice must discard twice as large a system as she does
here: $I(A;R)_\ph$ cbits per copy~\cite{GPW05}. In fact, it is
clear that we can derive that result from ours: after Alice's
unitary, the state remaining between $A_2$ and $R$ is almost a
product since Alice's entanglement with the reference gets
transferred to Bob, so Alice only needs to discard the system
$A_1$ of roughly $\frac{n}{2} I(A;R)$ qubits, which she can do by
erasing it entirely via random Pauli operations, at randomness
cost amounting to $I(A;R)$ cbits per copy.

\section{Fully quantum Slepian-Wolf: one-shot version} \label{sec:oneShotMother}

While the tensor power structure of $(\ket{\ph}^{ABR})^{\ox n}$
allows the fully quantum Slepian-Wolf inequality (\ref{eqn:FQSW})
to be expressed conveniently in terms of mutual information
quantities, our approach allows us to treat arbitrary input states
without such structure as well. In this section, we will prove a
general ``one-shot'' version of the fully quantum Slepian-Wolf
result that leads quickly to inequality (\ref{eqn:FQSW}) in the
special case where the input state is a tensor power.

For this section, we will therefore dispense with $\ket{\ph}^{\ox
n}$ and instead study a general state $\ket{\ps}^{ABR}$ shared
between Alice, Bob and the reference system. We also eliminate the
Schumacher compression step: assume that $A$ has been decomposed
into subsystems $A_1$ and $A_2$ satisfying $d_A = d_{A_1}
d_{A_2}$.

The following inequality is the one-shot version of
fully quantum Slepian-Wolf:

\begin{theorem}[One-shot, fully quantum Slepian-Wolf bound]
 \label{thm:trueoneShotMother}
There exist isometries $U^{A \rightarrow A_1 A_2}$ and $V^{A_1 B
\rightarrow \widehat{B} \tilde{B}}$
such that
$$
\Big\|( V \circ U) \psi^{RAB} ( V \circ U)^\dagger -
\psi^{R\widehat{B}} \otimes \Phi^{A_2 \tilde{B}} \Big\|_1 \leq
2\left[ \frac{2 d_A d_R}{d^2_{A_1}}
 \Big\{ \Tr[(\ps^{AR})^2] + 2 \Tr[(\ps^A)^2]\Tr[(\ps^R)^2] \Big\}
\right]^{1/4},
$$
where $W^{\widehat{B} \rightarrow AB} \ket{\psi}^{R\widehat{B}}  =
\ket{\psi}^{RAB}$ for some isometry $W$.
\end{theorem}

The protocol corresponding to the above theorem consists
of Alice performing $U$, sending the $A_1$ system to Bob,
and Bob performing $V$. The number of qubit channels used up is
$\log d_{A_1}$ whereas the number of ebits distilled is
$\log d_{A_2} = \log d_{A} - \log d_{A_1}$.

The main ingredient is the following decoupling theorem.

\begin{theorem}[Decoupling] \label{thm:oneShotMother}
Let $\s^{A_2 R}(U)
= \Tr_{A_1}[ (U \ox I^R) \ps^{AR} (U^\dg\ox I^R) ]$ be the state
remaining on $A_2 R$ after the unitary transformation $U$ has been
applied to $A = A_1 A_2$.
Then
\begin{multline} \label{eqn:oneShotMother}
\int_{\UU(A)} \Big\| \s^{A_2 R}(U) - \s^{A_2}(U) \ox \s^R(U)
\Big\|_1^2 \, dU \leq
 \frac{d_A d_R}{d_{A_1}^2}
 \Big\{ \Tr[(\ps^{AR})^2] + \Tr[(\ps^A)^2]\Tr[(\ps^R)^2] \Big\}.
\end{multline}
\end{theorem}
The theorem quantifies how distinguishable $\s^{A_2 R}(U)$ will be
from its completely decoupled counterpart $\s^{A_2}(U) \ox \s^R(U)$
if $U$ is chosen at random according to the Haar measure. As a first
observation, note that as $d_{A_1}$ grows, the two states become
progressively more indistinguishable. Also, the upper bound on the
right hand side is expressed entirely in terms of the dimensions of
the spaces involved and the purities $\Tr[(\ps^{AR})^2]$,
$\Tr[(\ps^A)^2]$ and $\Tr[(\ps^R)^2]$. In the tensor power source
setting, both dimensions and purities can be tightly bounded by functions
of the corresponding entropies, but in the one-shot setting they
must be distinguished.

In many situations of interest, the first term in the upper bound
dominates. In such cases, in order to assure a good approximation,
it suffices that
\begin{equation}
\log d_{A_1} \gg \frac{1}{2}\Big[ \log d_A + \log d_R + \log
\Tr[(\ps^{AR})^2] \Big].
\end{equation}
This expression plays the role of the $\smfrac{1}{2} I(A;R) =
\smfrac{1}{2}[ H(A) + H(R) - H(AR) ]$ from the FQSW resource
inequality (\ref{eqn:FQSW}) in the one-shot setting.

According to the proof strategy outlined in the previous section,
if $\s^{A_2 R}(U)$ is close to $\s^{A_2}(U) \ox \s^R(U)$, then
$\s^{A_2 R}(U)$ has a purification which is itself close to a
product state. This argument will be made quantitative in
the proof of Theorem \ref{thm:trueoneShotMother}.

The proof of the Theorem \ref{thm:oneShotMother} is quite
straightforward. We will evaluate the corresponding average over
the unitary group exactly for the Hilbert-Schmidt norm and then
use simple inequalities to extract inequality
(\ref{eqn:oneShotMother}). The evaluations of the relevant
averages are mechanical but slightly lengthy calculations. The
reader is advised that the proofs of lemmas \ref{lem:Schur},
\ref{lem:HSoneShotBound} and \ref{lem:whataboutA_2} are devoted
entirely to the calculation of such averages and can be skipped on
a first reading without impairing understanding of the rest of the
paper.

Before starting in earnest, we perform a calculation whose result
will be re-used several times. Recall from the notation summary in
the introduction that $F^{A_2 R}$ is the operator that swaps the
composite system $A_2 R$ with a duplicate composite system $A_2'
R'$, and that $\Pi^A_{+(-)}$ is the projector onto the
(anti-)symmetric subspace of $A$.
\begin{lemma} \label{lem:Schur}
\begin{equation}
\int_{\UU(A)} (U^\dg \ox U^\dg \ox I^{RR'})F^{A_2 R}(U \ox U \ox
I^{RR'}) \, dU
= [p \Pi^A_{+} + q \Pi^A_{-}]\ox F^R,
\end{equation}
where
\begin{equation} \label{eqn:pq}
p = \frac{d_{A_1} + d_{A_2}}{d_A + 1}
 \quad \mbox{and} \quad
 q = \frac{d_{A_1} - d_{A_2}}{d_A - 1}.
\end{equation}
\end{lemma}
\begin{proof}
Let $X$ be Hermitian. By Schur's Lemma,
\begin{equation}
\int_{\UU(A)} (U^\dg \ox U^\dg) X (U \ox U) \, dU
= \a_+(X) \Pi^A_{+} + \a_-(X) \Pi^A_{-},
\end{equation}
with the coefficients $\a_\pm(X) = \Tr(X
\Pi^A_\pm)/\rank(\Pi^A_\pm)$. Recall that $\Pi^A_\pm =
\smfrac{1}{2}(I^{AA'} \pm F^A)$.
\begin{eqnarray}
\rank(\Pi^A_\pm) \, \a_\pm(F^{A_2})
 &=& \Tr(\Pi^A_\pm F^{A_2}) \\
 &=& \frac{1}{2}\Tr\big[\big(I^{AA'}\pm F^{A_1} \ox F^{A_2}\big)F^{A_2}\big] \\
 &=& \frac{1}{2}\big[ \Tr(I^{A_1 A_1'} \ox F^{A_2}) \pm \Tr(F^{A_1}\ox I^{A_2
A_2'}) \big] \\
 &=& \frac{1}{2}[d_{A_1}^2 d_{A_2} \pm d_{A_1} d_{A_2}^2].
\end{eqnarray}
The second line uses the identity $F^A = F^{A_1} \ox F^{A_2}$. The
third follows from $F^2 = I$ and the explicit inclusion of
previously implicit identity operators to help in the evaluation
of the trace in line four. The formula then follows after a little
algebra, using that $F^{A_2 R} = F^{A_2} \ox F^R$ and
$\rank(\Pi^A_\pm) = d_A(d_A \pm 1)/2$.
\end{proof}

The next step is an exact evaluation of the Hilbert-Schmidt
analogue of the decoupling theorem.
\begin{lemma} \label{lem:HSoneShotBound}
\begin{multline} \label{eqn:HSoneShotBound}
\int_{\UU(A)} \Big\| \s^{A_2 R}(U) - \s^{A_2}(U) \ox \s^R(U)
\Big\|_2^2 \,
dU =\\
 \frac{d_{A_1}d_{A_2}^2 - d_{A_1}}{d_A^2 - 1}
 \Big\{ \Tr[(\ps^{AR})^2] -2\Tr[\ps^{AR}(\ps^A\ox\ps^R)] +
\Tr[(\ps^A)^2]\Tr[(\ps^R)^2]
 \Big\}.
 \quad \quad
\end{multline}
\end{lemma}
\begin{proof}
Note that
\begin{equation} \label{eqn:HSexpand}
\big\| \s^{A_2 R} - \s^{A_2} \ox \s^R \big\|_2^2
 = \Tr[(\s^{A_2 R})^2] - 2 \Tr[\s^{A_2 R}(\s^{A_2}\ox \s^R)]
 + \Tr[(\s^{A_2})^2]\Tr[(\s^R)^2].
\end{equation}
Starting with the first term,
\begin{eqnarray} \label{eqn:mainInt}
\int_{\UU(A)} \Tr[(\s^{A_2 R}(U))^2]\, dU
 &=& \int \Tr\big[\big(\s^{A_2 R}(U) \ox \s^{A_2' R'}(U)\big) F^{A_2R}\big] \,dU
\\
 &=& \int \Tr\big[ \big( \Tr_{A_1}(U \ps^{AR} U^\dg)
    \ox \Tr_{A_1'}(U \ps^{A'R'} U^\dg)\big) F^{A_2R}\big] \, dU \\
 &=& \Tr\big[ (\ps^{AR} \ox \ps^{A'R'} ) \cdot \int (U^\dg \ox U^\dg)
    (I^{A_1A_1'} \ox F^{A_2R})
    (U \ox U)  \,dU \big] \\
 &=& \Tr\big[ (\ps^{AR} \ox \ps^{A'R'} ) \cdot (p \Pi^A_{+} + q \Pi^A_{-})\ox
F^R \big]\\
 &=& \frac{p+q}{2} \Tr[(\ps^R)^2] + \frac{p-q}{2} \Tr[(\ps^{AR})^2],
\end{eqnarray}
where $p$ and $q$ are defined as in Eq.~(\ref{eqn:pq}). In the
fourth line we've used the result of Lemma~\ref{lem:Schur}, and in
the fifth the identity $\Pi^A_\pm = \smfrac{1}{2}(I^{AA'}\pm
F^A)$. The third term in Eq.~(\ref{eqn:HSexpand}) can also be
evaluated using this formula and the observation that $\s^R(U) =
\ps^R$, giving
\begin{equation}
\int_{\UU(A)} \Tr[(\s^{A_2})^2]\Tr[(\s^R)^2] \, dU =
 \Big\{ \frac{p+q}{2} +\frac{p-q}{2} \Tr[(\ps^{A})^2] \Big\}
    \Tr[(\ps^R)^2].
\end{equation}
That leaves the second term of Eq.~(\ref{eqn:HSexpand}), which can
be calculated in the same way as Eq.~(\ref{eqn:mainInt}), with the
result that
\begin{equation}
\int_{\UU(A)} \Tr[\s^{A_2 R}(\s^{A_2}\ox \s^R)] \, dU
 = \frac{p+q}{2} \Tr[(\ps^R)^2] + \frac{p-q}{2} \Tr[\ps^{AR}(\ps^A \ox
 \ps^R) ].
\end{equation}
Substituting back into Eq.~(\ref{eqn:HSexpand}) shows that
$\int_{\UU(A)} \| \s^{A_2 R}(U) - \s^{A_2}(U) \ox \s^R(U) \|_2^2
\, dU$ is equal to
\begin{equation}
\frac{p-q}{2}\Big\{ \Tr[(\ps^{AR})^2] - 2
\Tr[\ps^{AR}(\ps^A\ox\ps^R)] + \Tr[(\ps^A)^2]\Tr[(\ps^R)^2]
\Big\},
\end{equation}
which, after substitution for $p$ and $q$, yields
(\ref{eqn:HSoneShotBound}).
\end{proof}
The decoupling theorem is now an easy
corollary:

\begin{proof}\hspace{-1.5mm}{\bf of Theorem \ref{thm:oneShotMother}:}
The Cauchy-Schwarz inequality can be used to relate the two norms:
$\| \cdot \|_1^2 \leq d_{A_2} d_R \| \cdot \|_2^2$. Also,
$\Tr[\ps^{AR}(\ps^A\ox\ps^R)]$ is nonnegative. Finally,
\begin{equation} \label{eqn:actuallyItsSimple}
\frac{d_{A_1} d_{A_2}^2 - d_{A_1}}{d_A^2-1}
 \leq \frac{1}{d_{A_1}}
\end{equation}
holds for all $d_{A_1} \geq 1$.
\end{proof}

To make contact with the Theorem \ref{thm:trueoneShotMother}, we
must verify that the state Bob  shares with Alice is close to
maximally entangled. This is true only if $ \s^{A_2}(U)$ is almost
maximally mixed.

\begin{lemma} \label{lem:whataboutA_2}
 \label{lemma:maxmix}
$$
 \int_{\UU(A)} \Big\| \s^{A_2}(U) - \frac{I^{A_2}}{d_{A_2}} \Big\|_1^2 \,
 dU \leq \frac{d_{A}}{d_{A_1}^2} \Tr[(\ps^A)^2].
$$
\end{lemma}

\begin{proof}
\begin{eqnarray}
 \int_{\UU(A)} \Big\| \s^{A_2}(U) - \frac{I^{A_2}}{d_{A_2}} \Big\|_1^2 \,
 dU
 &\leq& d_{A_2} \int_{\UU(A)} \Big\| \s^{A_2}(U) - \frac{I^{A_2}}{d_{A_2}}
        \Big\|_2^2 \, dU \\
 &=& d_{A_2} \int_{\UU(A)} \Tr[(\s^{A_2}(U))^2] dU - 1 \\
 &=& d_{A_2} \Big\{ \frac{p+q}{2} + \frac{p-q}{2} \Tr[(\ps^{A})^2]
    \Big\} - 1 \\
 &\leq& d_{A_2} \frac{p-q}{2} \Tr[(\ps^{A})^2] \\
 &\leq& \frac{d_{A}}{d_{A_1}^2} \Tr[(\ps^A)^2].
 \label{eqn:ABentanglement}
\end{eqnarray}
The first line is Cauchy-Schwarz, the third is an integral that
we already performed in proving Lemma \ref{lem:HSoneShotBound},
the fourth is by $\frac{p+q}{2} \leq \frac{1}{d_{A_2}}$,
and the last is an application of
Eq.~(\ref{eqn:actuallyItsSimple}).
\end{proof}

We are now ready to prove Theorem \ref{thm:trueoneShotMother}.

\begin{proof}\hspace{-1mm}{\bf of Theorem \ref{thm:trueoneShotMother}:}
\begin{equation}
 \label{eqn:ABentanglement2}
\begin{split}
& \int_{\UU(A)} \Big\| \s^{A_2R}(U) - \frac{I^{A_2}}{d_{A_2}}
\otimes  \s^{R}(U)\Big\|_1^2 \,
 dU \\
 &\leq  \int_{\UU(A)}
\Big( \Big\| \s^{A_2R}(U) -  \s^{A_2}(U) \otimes  \s^{R}(U)
\Big\|_1
 + \Big\| \s^{A_2}(U) - \frac{I^{A_2}}{d_{A_2}}   \Big\|_1
\Big)^2\,
 dU \\
 &\leq 2 \int_{\UU(A)}
\Big\| \s^{A_2R}(U) -  \s^{A_2}(U) \otimes  \s^{R}(U) \Big\|_1^2
 + \Big\| \s^{A_2}(U) - \frac{I^{A_2}}{d_{A_2}}   \Big\|_1^2 \,
 dU \\
 &\leq \frac{2 d_A d_R}{d^2_{A_1}}
 \Big\{ \Tr[(\ps^{AR})^2] + 2 \Tr[(\ps^A)^2]\Tr[(\ps^R)^2] \Big\} .
\end{split}
\end{equation}
The first line is the triangle inequality,
the second is the Cauchy-Schwartz inequality,
and the third follows from Theorem \ref{thm:oneShotMother},
Lemma \ref{lemma:maxmix} and $d_{R} \Tr[(\ps^R)^2] \geq 1$.
Observe that there exists a particular $U$
such that
$\Big\| \s^{A_2R} - \frac{I^{A_2}}{d_{A_2}} \otimes  \s^{R} \Big\|_1^2$
is bounded as in (\ref{eqn:ABentanglement2}).

The final ingredient is Uhlmann's theorem~\cite{U76}, in the
version of Lemma 2.2 of~\cite{DHW05}: If  $\| \rho^C - \sigma^C
\|_1 \leq \epsilon$, $\rho^{BC}$ is a purification of $\rho^{C}$,
and $\s^{DC}$ is a purification of $\s^{C}$ then there exists an
isometry $V^{B \rightarrow D}$ such that $\| (V^B \otimes I^C)
\rho^{BC}  (V^B \otimes I^C)^\dagger
 - \sigma^{BC} \|_1 \leq 2 \sqrt{\epsilon}$.
Since $\Phi^{A_2 \tilde{B}} \otimes \psi^{R\widehat{B}}$ is a
purification of $\frac{I^{A_2}}{d_{A_2}} \otimes  \s^{R}$ and $U
\psi^{RAB} U^\dagger$ is a purification of
 $\s^{A_2R}$, there is an isometry
$V^{A_1 B \rightarrow \tilde{B} \widehat{B}}$ such that the
statement of the theorem holds.

\end{proof}

\section{Father from FQSW: one-shot version}
\label{sec:father_1}

A few simple observations will allow us to transform the one-shot
FQSW protocol into a one-shot father protocol. The father
implements entanglement-assisted noiseless quantum communication
over a noisy channel $\cN^{A\rar B}$. The protocol consumes
entanglement initally shared between Alice and Bob in registers we
will call $A_3$ and $B_3$. Mathematically, we verify that the
protocol implements noiseless quantum communication by applying it
to one half of a maximally entangled state, the other half of
which is held by a reference system $R$. This is equivalent to
verifying that after the application of $\cN^{A\rar B}$, the
reference system $R$ is decoupled from the channel's environment
$E$. In the one-shot FQSW protocol, the objective was to decouple
$R$ and $A_2$.

We make the corresponding replacements in Theorem
\ref{thm:trueoneShotMother}:
\begin{center}
\begin{tabular}{|c|c|}
 \hline
 \, Father \, & \, FQSW \, \\
 \hline \hline
  $B_3$  & $A_1$ \\ \hline
  $R$    & $A_2$ \\ \hline
  $B_3 R$    & $A$ \\ \hline
  $E$    & $R$ \\ \hline
\end{tabular}
\end{center}
\begin{figure}
 \begin{center}
  \includegraphics[width=5in,bbllx=8,bblly=400,bburx=784,bbury=620]{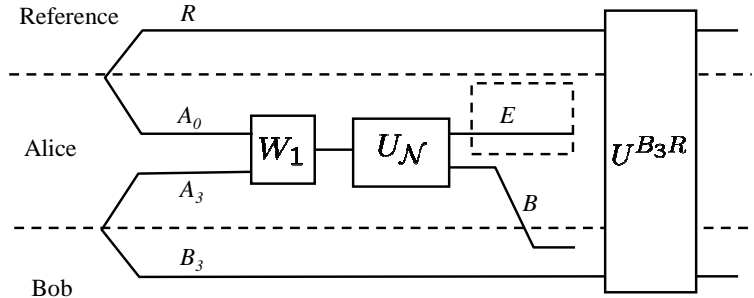}
 \end{center}
 \caption{Partial reduction from the father to the mother. Dotted lines are used to demarcate domains controlled
 by the different partipicants and solid lines represent quantum
 information. Note that Alice starts the protocol sharing one
 maximally entangled state with the reference, $\ket{\Phi_0}^{A_0
 R}$, and another with Bob, $\ket{\Phi_3}^{A_3 B_3}$.
 The unitary transformation $U^{B_3 R}$ comes from
 an application of the FQSW theorem with $B_3 R$ replacing $A_1 A_2$. After the application of the unitary,
 the registers $R$ and $E$ are nearly decoupled, as desired, but unfortunately, because it requires acting
 on the reference system $R$, $U^{B_3 R}$ cannot be used in this way.}
 \label{fig:father1}
\end{figure}
Thus, there exist isometries $U^{B_3 R \rightarrow B_3 R}$ and
$V^{B_3 B \rightarrow \widehat{B} \tilde{B}}$
such that
$$
\|( V \circ U) \psi^{B_3RBE} ( V \circ U)^\dagger -
\psi^{\widehat{B}E} \otimes \Phi_0^{R \tilde{B}} \|_1 \leq
2\!\left[ \frac{2 d_{B_3 R} d_E}{d^2_{B_3}}
 \Big\{ \!\!\Tr[(\ps^{B_3RE})^2] + 2 \Tr[(\ps^{B_3R})^2]\Tr[(\ps^E)^2] \Big\}
\right]^{1/4} \!\!\!\!,
$$
where $W^{\widehat{B} \rightarrow B_3RB} \ket{\psi}^{\widehat{B}E}
= \ket{\psi}^{B_3RBE}$ for some isometry $W$. In particular, let
\begin{equation}
\label{partic}
\ket{\psi}^{B_3RBE} = U_\cN^{A \rightarrow BE}
\circ W_1^{A_0 A_3 \rightarrow A} (\ket{\Phi_0}^{R A_0}
\ket{\Phi_3}^{B_3 A_3})
\end{equation}
for some Stinespring dilation $U_\cN^{A \rightarrow BE}$ of a
noisy channel $\cN^{A \rightarrow BE}$, and isometry $W_1^{A_0 A_3
\rightarrow A}$. Since $V^{B_3 B\rar \widehat{B}\tilde{B}}$ acts
entirely on systems held by Bob, it could be performed by him as a
decoding operation. The isometry $U^{B_3 R \rar B_3 R}$, on the
other hand, acts on the reference system, which is not allowed to
participate actively in the protocol. The situation up to this
point is depicted in Figure \ref{fig:father1}.
However, because $\Phi_0^{RA_0} \otimes \Phi_3^{B_3 A_3}$ is
maximally entangled between $A_3A_0$ and $B_3R$,
$$
U^{B_3 R \rightarrow B_3 R} (\ket{\Phi_0}^{R A_0}
\ket{\Phi_3}^{B_3 A_3})
= (U^T)^{A_3 A_0 \rightarrow A_3 A_0} (\ket{\Phi_0}^{R A_0}
\ket{\Phi_3}^{B_3 A_3}),
$$
where $T$ denotes transposition. Thus, the effect of $U$ can be
achieved by acting instead with $U^T$ on $A_3 A_0$, systems held
by Alice. Defining $W_2^{A_0 A_3 \rightarrow A}
= W_1 \circ U^T$, we get
\begin{equation}
\begin{split}
&\|( V \circ U_\cN \circ W_2) ( \Phi_0^{R A_0}
\otimes \Phi_3^{B_3 A_3}) ( V \circ U_\cN \circ W_2)^\dagger -
\psi^{\widehat{B}E} \otimes \Phi_0^{R \tilde{B}} \|_1 \\
&\phantom{=====}
  \leq  2\left[ \frac{2 d_{A_0 A_3} d_E}{d^2_{A_3}}
 \Big\{ \Tr[(\ps^{B})^2] + \frac{ 2}{ d_{A_0 A_3}}\Tr[(\ps^E)^2] \Big\}
\right]^{1/4}.
\end{split}
\end{equation}
This is precisely the setting of the father protocol, as
illustrated in Figure~\ref{fig:father2}.

Alice needs to transfer the purification of some maximally mixed
state $\Phi_0^{R}$ to Bob. The resources at their disposal are the
channel $\cN^{A \rightarrow B}$ and a maximally entangled state
$\Phi_3^{B_3 A_3}$. Alice performs the encoding $W_2$, sends the
resulting state through the channel $\cN$ and Bob decodes with
$V$. The number of ebits used up is $\log d_{A_3}$ whereas the
number of qubits sent is $\log d_{A_0}$.
\begin{figure}
 \begin{center}
  \includegraphics[width=5in,bbllx=8,bblly=380,bburx=784,bbury=640,clip=true]{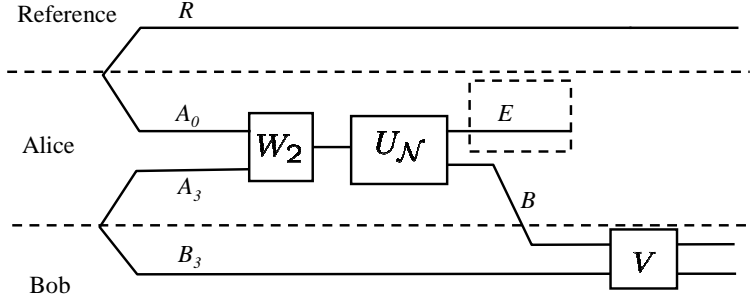}
 \end{center}
 \caption{Final version of the father protocol generated from FQSW.
 As Figure \ref{fig:father1} makes clear, $U^{B_3 B}$ was required to act on
 one half of a maximally entangled state, the other half of which found in $A_3 A_0$, register held by
 Alice. Thus, Alice can instead implement the encoding operation $W_2 = W_1 \circ U^T$. Bob performs
 the decoding operation $V$ mandated by FQSW, resulting in the one-shot father.}
 \label{fig:father2}
\end{figure}

\section{Fully quantum reverse Shannon theorem: one-shot version}
\label{sec:qrst-oneshot}
The quantum reverse Shannon theorem was conceived of in~\cite{BSST99,BSST02},
and is proved in full in~\cite{BDHSW06}. It asserts that in the presence of
entanglement, a noisy quantum channel $\cN$ can be simulated by
$C_E(\cN)$ cbits of forward classical communication per copy of the
channel, where $C_E$ is the entanglement-assisted capacity of the channel.

Here, following~\cite{D05b}, we demonstrate how, by running the
mother protocol backwards, one obtains a simple proof of a fully
quantum version of this result. The Stinespring dilation
$U_{\cN}:A' \rightarrow BE$ of $\cN^{A' \rar B}$ is simulated in
such a way that $E$ ends up with Alice. For that reason, we say
that the protocol simulates the \emph{feedback channel} associated
to $\cN^{A' \rar B}$.

\begin{figure}
 \begin{center}
  \includegraphics[width=5in,bbllx=8,bblly=430,bburx=784,bbury=620]{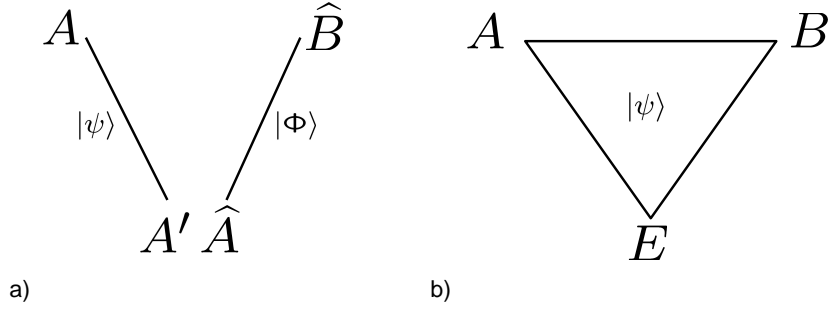}
 \end{center}
 \caption{a) The starting point for FQRS, a pair of pure entangled
 states. The system $A$ is a purification of Alice's input system $A'$ while
 $\tilde{A}\tilde{B}$ holds the entanglement that Alice-Bob will consume to
 execute the protocol.
 b) After execution of the protocol, the reference system $A$ is unchanged while
 Alice receives the environment feedback system $E$ and Bob receives his share $B$
 of the state $\ket{\psi}^{ABE} = U_{\cN}^{A'\rar BE} \ket{\psi}^{AA'}$.}
 \label{fig:FQRS}
\end{figure}

Ultimately, in section \ref{sec:qrst-iid}, we will show the
 \emph{fully quantum reverse Shannon} (FQRS)
resource inequality
\begin{equation}
 \frac{1}{2}I(A;B)_\varphi [q\rightarrow q]
     + \frac{1}{2}I(B;E)_\varphi [qq]
  \geq \left\langle U_{\cN}^{A'\rightarrow BE} : \rho^{A'} \right\rangle,
\label{eqn:FQRS}
\end{equation}
where $\ket{\varphi}^{ABE} = U_{\cN}^{A'\rightarrow BE} \ket{\varphi}^{AA'}$
 and
$\ket{\varphi}^{AA'}$ is a purification of $\rho^{A'}$. In this
section, we will actually prove a one-shot version of this
resource inequality, by a simple re-interpretation of the systems
of the mother, and running her backwards in time. The task is to
simulate with high fidelity the feedback channel $U_{\cN}:A'
\rightarrow B E$ on a source ${\psi}^{AA'}$, using some maximal
entanglement $\Phi^{\tilde{A}\tilde{B}}$ and quantum communication
of a system $A_1$ of dimension $d_{A_1}$. From a mathematical
point of view, the state $\ket{\psi}^{ABE} = U_{\cN}^{A'\rar BE}
\ket{\psi}^{AA'}$ has to be created from $\ket{\psi}^{AA'} \ox
\ket{\Phi}^{\tilde{A}\tilde{B}}$, as illustrated in
Figure~\ref{fig:FQRS}.

Recall that the one-shot FQSW protocol created a product state
starting from an arbitrary pure tripartite entangled state,
whereas here the goal is to do the reverse. Hence the need to run
the protocol backwards in time. To help see the appropriate choice
of relabellings, note that in the FQSW case, Bob holds
purifications of the $R$ and $A_2$ systems, called $\widehat{B}$
and $\tilde{B}$ respectively. In the present setting, Alice starts
holding purifications $A'$ and $\tilde{A}$ of $A$ and $\tilde{B}$
respectively. Matching the corresponding systems suggests the
following replacements in the one-shot mother:
\begin{center}
\begin{tabular}{|c|c|}
 \hline
 \, FQRS \, & \, FQSW \, \\
 \hline \hline
  $A'$       & $\widehat{B}$ \\ \hline
  $A$        & $R$ \\ \hline
  $B$        & $A$ \\ \hline
  $E$        & $B$ \\ \hline
  $\tilde{A}$ & $\tilde{B}$ \\ \hline
  $\tilde{B}$ & $A_2$ \\ \hline
\end{tabular}
\end{center}
A comparison of Figure \ref{fig:FQRS} with the FQRS analogue,
Figure \ref{fig:FQSW} is also very helpful for clarifying the role
of the substitutions. We can interpret
theorem~\ref{thm:trueoneShotMother} as saying that there exist
isometries $U^{B \rightarrow A_1 \tilde{B}}$ and $V^{A_1 E
\rightarrow A' \tilde{A}}$ such that
\[\begin{split}
  \bigl\| \psi^{ABE} -
             (U^\dagger \circ V^\dagger)
                 (\psi^{AA'} \otimes \Phi^{\tilde{A}\tilde{B}})
             (U^\dagger \circ V^\dagger)^\dagger \bigr \|_1
   &=     \|( V \circ U) \psi^{ABE} ( V \circ U)^\dagger -
             \psi^{AA'} \otimes \Phi^{\tilde{A}\tilde{B}} \|_1  \\
   &\!\!\!\!\!\!\!\!\!
    \leq  2\!\left[ \frac{2 d_B d_A}{d^2_{A_1}}
                    \Big\{ \Tr[(\ps^{AB})^2] + 2 \Tr[(\ps^A)^2]\Tr[(\ps^B)^2]
\Big\}
             \right]^{1/4} \!\!.
\end{split}\]
In other words, Alice performs $V^\dagger : A'\tilde{A} \rar A_1E$
on her part of the system, and sends $A_1$ to Bob; she keeps $E$
which will be the environment of the channel. (Note that because
of the input state, $V^\dg$ is actually a well-defined isometry!)
Bob can perform the isometry $U^\dagger : A_1\tilde{B} \rar B$ to
obtain the channel output in $B$.

\section{Fully quantum Slepian-Wolf: i.i.d. version}
 \label{sec:iid}
We return now to the setting where Alice, Bob and the reference
system share the state $\ket{\ps'} = (\ket{\ph}^{ABR})^{\ox n}$.
This is often called the i.i.d. case because each copy of the
state is identical and independently distributed. Combining the
one-shot, fully quantum Slepian-Wolf result with Schumacher
compression will lead to the FQSW resource inequality
(\ref{eqn:FQSW}).
In Appendix A we show the following: For any $\epsilon, \delta >
0$ and sufficiently large $n$, we can define projectors $\Pi_A,
\Pi_B, \Pi_R$ onto the $\delta$-typical subspaces of the systems
indicated by the subscripts such that the following properties
hold for any subsystem $F = A, B, R$:
\begin{enumerate}[i)]
\item $\| \cE(\ps') - \ps' \|_1 \leq \epsilon$,
\item $\| \ps - \ps' \|_1 \leq \epsilon$,
\item $ 2^{n[H(F) - \delta]}   \leq  \rank \Pi_F \leq 2^{n[H(F)+ \delta]}$,
\item $ \Tr[(\ps^F)^2] \leq 2^{-n[H(F) - \delta]}$.
\end{enumerate}
Here $\cE^{A \rightarrow A^{typ}}$ is the Schumacher compression operation
(one of whose Kraus elements is $\Pi_A$) and
$\ket{\ps}$ the normalized
version of the state
\begin{equation}
(\Pi_A \ox \Pi_B \ox \Pi_R) \ket{\ps'}.
\end{equation}

While we are concerned with the output of the protocol when it is
applied to the state $\ket{\ps'}=(\ket{\ph}^{ABR})^{\ox n}$, by
properties i) and ii) we can analyze its effect on the nearly
indistinguishable $\ket{\ps}$ instead.

Thanks to the properties of the typical projectors, namely
properties iii) and iv), the various quantities appearing in the
upper bound of Theorem \ref{thm:oneShotMother} get replaced by
entropic formulas in the i.i.d. case. For an arbitrary subsystem
$F$, let $F^{typ}$ denote the support of $\Pi_F$ and assume
$A^{typ} = A_1 \ox A_2$. By Theorem \ref{thm:trueoneShotMother},
there exist isometries $U^{A^{typ} \rightarrow A_1 A_2}$ and
$V^{A_1 B \rightarrow \widehat{B} \tilde{B}}$ such that
\begin{equation}
\label{skoro}
\begin{split}
& \big\|( V \circ U) \psi^{R^{typ}A^{typ}B} ( V \circ U)^\dagger -
\psi^{R^{typ}\widehat{B}} \otimes \Phi^{A_2 \tilde{B}} \big\|_1\\
&\phantom{====}
 \leq  2\left[ \frac{2 d_{A^{typ}} d_{R^{typ}}}{d^2_{A_1}}
 \Big\{ \Tr[(\ps^{A^{typ}R^{typ}})^2] + 2 \Tr[(\ps^{A^{typ}})^2]
   \Tr[(\ps^{R^{typ}})^2] \Big\} \right]^{1/4} \\
&\phantom{====}
 \leq 2(4 \cdot 2^{n [I(A;R) + 3 \delta]}/d^2_{A_1})^{1/4}.
\end{split}
\end{equation}
Choosing  $\log d_{A_1} = n [ I(A;R)/2 + 2\d ]$, the bound of
Eq.~(\ref{skoro}) becomes less than or equal to $\sqrt{8} 2^{-n
\d/4}$.

Since $\psi$, $\cE(\psi')$ and $\psi'$ are close, performing the protocol
on the Schumacher compressed state $\cE(\psi')$  will also do well.
More precisely, a double application of the
triangle inequality and properties i) and ii) give
$$
 \|( V \circ U) \cE ({\psi'}^{RAB}) ( V \circ U)^\dagger -
{\psi'}^{R\widehat{B}} \otimes \Phi^{A_2 \tilde{B}} \|_1 \leq 2
\epsilon +  \sqrt{8} 2^{-n \d/4}.
$$
The number of qubit channels used up is thus
$n [ I(A;R)/2 + 2\d ]$, whereas the number of ebits distilled is
$\log d_{A_2} = \log d_{A^{typ}} - \log d_{A_1} \geq
n[I(A;B)/2 -  3\d]$.

%

\section{Father: i.i.d. version} \label{sec:father:iid}

In the i.i.d. father setting described by the resource inequality
(\ref{eqn:father}), Alice and Bob are given a channel of the form
$(\cN^{A' \rightarrow B})^{\otimes n}$. Choose a Stinespring
dilation ${U_\cN}^{A' \rightarrow BE}$ such that $\cN(\rho) =
\Tr_E U \r U^\dg$ and define $\ket{\ph}^{ABE}
= {U_\cN}\ket{\ph}^{AA'}$.
Let $\ket{\ps}$ and $\ket{\ps'}$ be as in the previous section,
only with $R$ replaced by $E$. Now define
$\Pi^t_A$ to be the projector onto a particular \emph{typical}
type $t$ and define $\ket{\psi_t'}$ and $\ket{\psi_t}$ to be the
 normalized versions of the states  ${\Pi^t_A \ket{\psi'}}$
and $\Pi^t_A \ket{\psi}$, respectively.
In Appendix A it is shown that there exists a particular $\Pi^t_A$
such that the following properties hold:
\begin{enumerate}[i)]
\item $ \ps_t^A = I/ (\rank \Pi^t_A)$ ,
\item $\| \ps_t - \ps_t' \|_1 \leq \epsilon$,
\item $ 2^{n[H(F) - \delta]}   \leq  \rank \Pi_F \leq 2^{n[H(F)+ \delta]}$,
\item $ \Tr[(\ps_t^F)^2] \leq 2^{-n[H(F) - \delta]}$.
\item $ 2^{n[H(A) - \delta]}   \leq  \rank \Pi^t_A \leq 2^{n[H(A)+ \delta]}$
\end{enumerate}
Let $A_{t}$ denote the support of $\Pi^t_A$.
By property i), $\ket{\psi'_t}^{A_tBE}$ is the result of
sending a maximally entangled state proportional to
$\ket{\Phi}^{A_t A'_t}
= (\Pi_t^{A} \otimes \Pi_t^{A'}) (\ket{\ph}^{AA'})^{\otimes n}$
through ${U_\cN}^{\otimes n}$.
Similarly, $\ket{\psi_t}^{A_tB^{typ}E^{typ}}$ arises from the
modified channel $(\Pi_B \otimes \Pi_E) \circ {U_\cN}^{\otimes n}$.
Thus $\ket{\psi_t}^{A_tBE}$ is of the form $(\ref{partic})$,
and we can apply the results of section \ref{sec:father_1}.
Proceeding as in the previous section and using the above properties
we conclude that there exist  isometries $W_2^{A_0 A_3 \rightarrow
A}$ and  $V^{B_3 B \rightarrow \widehat{B} \tilde{B}}$ such that
\[
  \|( V \circ U_\cN^{\otimes n} \circ W_2) ( \Phi_0^{R A_0}
     \otimes \Phi_3^{B_3 A_3}) ( V \circ U_\cN^{\otimes n} \circ W_2)^\dagger -
      \psi_t^{\widehat{B}E} \otimes \Phi_0^{R \tilde{B}} \|_1 \\
  \leq 2 \epsilon +  \sqrt{8} 2^{-n \d/4}.
\]
The number of ebits used up is $\log d_{A_3} = n [ I(A;E)/2 + 2\d
]$ and the number of qubits transmitted is $\log d_{A_0} = \log
d_{A_{t}} - \log d_{A_3} \geq  n [ I(A;B)/2 - 3\d ]$, leading to
the asymptotic rates required by the father resource inequality.

\section{Fully quantum reverse Shannon theorem: i.i.d. version}
\label{sec:qrst-iid}

As in the previous two sections, we can consider the special case
in which  Alice and Bob want to simulate many realizations of the
channel $\cN:A\rightarrow B$, or rather its feedback isometry
$U_{\cN}:A\rightarrow BE$, relative to a source $\rho^A$. The FQRS
resource inequality~(\ref{eqn:FQRS}) was described in
section~\ref{sec:qrst-oneshot}. Just as in section \ref{sec:iid},
the resource inequality is achieved by mentally truncating the
state $\bigl(\ket{\ph}^{ABE}\bigr)^{\ox n}$ to its typical part,
introducing small disturbances, and then running the one-shot
protocol on the truncated state. We omit the details.

\section{Correlated source coding: distributed compression}
\label{sec:distributed}

One of the major applications of the state merging inequality
(\ref{eqn:merging}) is to the problem of distributed compression
with free forward (or indeed completely unrestricted) classical
communication. For this problem, Horodecki, Oppenheim and Winter
demonstrated that the resulting region of achievable rates has the
same form as the classical Slepian-Wolf problem~\cite{SW71,HOW05}.
In this section, we consider the application of the fully quantum
Slepian-Wolf inequality to distributed compression without
classical communication.

Because distributed compression studies multiple senders, it no
longer fits into the resource inequality framework as laid out
in~\cite{DHW05}. We therefore begin with some definitions
describing the task to be performed. A source provides Alice and
Bob with the $A$ and $B$ parts of a quantum state $\ket{\ps} =
(\ket{\ph}^{ABR})^{\ox n}$ purified by a reference system $R$.
They must independently compress their shares and transmit them to
a receiver Charlie. That is, they will perform encoding operations
$E_A$ and $E_B$ described by completely positive, trace-preserving
(CPTP) maps with outputs on systems $C_A$ and $C_B$ of dimensions
$2^{nQ_A}$ and $2^{nQ_B}$, respectively. The receiver, Charlie,
will then perform a decoding operation, again described by a CPTP
map, this time with output systems $\widehat{A}$ and $\widehat{B}$
isomorphic to $A^n$ and $B^n$. A rate pair $(Q_A,Q_B)$ will be
said to be achievable if for all $\e>0$ there exists an $N(\e) >
0$, such that for all $n \geq N(\e)$ there exists a corresponding
$(E_A,E_B,D)$ such that
\begin{equation}
\bra{\ps}^{R^n\widehat{A}\widehat{B}}
    (D \circ (E_A \ox E_B))(\ps^{R^nA^nB^n})
    \ket{\ps}^{R^n\widehat{A}\widehat{B}} \geq 1 - \e.
\end{equation}
The achievable rate region $\region(\ph)$ for a given $\ket{\ph}$
is the closure of the set of achievable rates. By time-sharing it
is a convex set.

The fully quantum Slepian-Wolf inequality provides a natural class
of protocols for this task. One party, say Bob, first Schumacher
compresses his share and sends it to Charlie. This is possible
provided $Q_B > H(B)_\ph$. The other party, in this case Alice,
then implements the fully quantum Slepian-Wolf protocol with
Charlie playing the role of Bob. This is possible provided $Q_A >
I(A;R)/2$. Looking at the total number of qubits required gives a
curious symmetrical formula:
\begin{equation}
Q_A + Q_B
 > \frac{1}{2} I(A;R)_\ph + H(B)_\ph
 = \frac{H(A)_\ph + H(B)_\ph + H(AB)_\ph}{2}
 =: \frac{1}{2} J(A;B)_\ph,
\end{equation}
introducing a new symbol $J(A;B) = H(A)+H(B)+H(AB)$ for the
characteristic rate sum above, a kind of quasi-mutual
information with a plus sign instead of minus.

By switching the roles played by Alice and Bob and also
time-sharing between the resulting two protocols, we find
\begin{theorem}
  \label{thm:distrib-inner}
  The region defined by
  \begin{align} \label{eqn:region}
    Q_A       &\geq \frac{1}{2} I(A;R)_\ph \nonumber\\
    Q_B       &\geq \frac{1}{2} I(B;R)_\ph \\
    Q_A + Q_B &\geq \frac{1}{2} J(A;B)_\ph \nonumber
  \end{align}
  is contained in the achievable rate region $\region(\ph)$.
  \qed
\end{theorem}

In fact, the region of Theorem~\ref{thm:distrib-inner} is in some
cases \emph{equal} to $\region(\ph)$, as we will see by proving a
general outer bound on the achievable rate region. Assume that
$(Q_A,Q_B) \in \region(\ph)$. To begin, fix $n > N(\e)$ and let
$W_A$ and $W_B$ be the environments for the Stinespring dilations
of the encoding operations $E_A$ and $E_B$. We may without loss of
generality assume that their dimensions $d_{W_A}$, $d_{W_B}$ are
bounded above by $d_A^{2n}$, $d_B^{2n}$, respectively, because
every CPTP map from a space of dimension $d$ to a space of
dimension at most $d$ can be written using at most $d^2$ Kraus
operators.

To bound $Q_A$, assume that Charlie has received both $C_B$
\emph{and} $W_B$, that is, all of $B^n$. Let $W_C$ be the
environment for the dilation of Charlie's $D$. Again, without loss
of generality we can assume that the Stinespring dilations are
implemented by preparing the environment systems in pure
unentangled states and then applying unitary transformations.
Because at the end of the protocol Charlie must have essentially
$A^nB^n$, which purifies $R^n$, the registers $W_AW_C$ have to be
in a pure state of their own, product with $R^n$ and Charlie's
output $\widehat{A}\widehat{B}$. Of course, this is not exactly
true, only with high fidelity, so we proceed to make these
statements rigorous.

Let $\ket{\xi}^{R^n\widehat{A}\widehat{B}W_A W_C}$ be the final
state after the application of the Stinespring dilations of the
encoding and decoding. By the fidelity condition,
\begin{equation*}
\lambda_{\max}(\xi^{R^n\widehat{A}\widehat{B}})
       \geq \Tr \bigl[\xi^{R^n\widehat{A}\widehat{B}}
\proj{\psi}^{R^n\widehat{A}\widehat{B}}\bigr]
       \geq 1-\epsilon,
\end{equation*}
where $\lambda_{\max}(\xi^{R^n\widehat{A}\widehat{B}})$ denotes
the maximum eigenvalue of $\xi^{R^n\widehat{A}\widehat{B}}$.
Therefore, $\ket{\xi}^{R^n\widehat{A}\widehat{B}W_AW_C}$ has
Schmidt decomposition
\begin{equation}
  \ket{\xi}^{R^n\widehat{A}\widehat{B}W_AW_C}
=\sum_i\sqrt{\lambda_i}\ket{v_i}^{R^n\widehat{A}\widehat{B}}\ket{w_i}^{W_AW_C},
\end{equation}
where $\lambda_1 = \lambda_{\max}\geq 1-\epsilon$, and consequently,
\begin{equation*}
  \Tr\bigl[ \proj{\xi}^{R_n\widehat{A}\widehat{B}W_AW_C}
            (\xi^{R^n\widehat{A}\widehat{B}}\otimes\xi^{W_AW_C})
     \bigr]
  \geq \sqrt{1-\epsilon}^2(1-\epsilon)^2 \geq 1-3\epsilon.
\end{equation*}
So, since the above is the fidelity between states,
\begin{equation*}
  \left\| \ket{\xi}^{R^n\widehat{A}\widehat{B}W_AW_C}
            - \xi^{R^n\widehat{A}\widehat{B}} \ox \xi^{W_AW_C} \right\|_1
  \leq 2\sqrt{3\epsilon},
\end{equation*}
by~\cite{FvG99}, and with the contractivity of the trace distance we now have
\begin{equation}
  \label{eq:almostproduct}
  \left\| \xi^{R^nW_A} - \xi^{R^n} \ox \xi^{W_A} \right\|_1
     \leq 2\sqrt{3\epsilon},
\end{equation}
We can now apply the Fannes inequality~\cite{F73} to yield:
\begin{equation}\begin{split}
  \label{eqn_1}
  \left| H(\xi^{R^nW_A}) - H(\xi^{R^n}\otimes\xi^{W_A}) \right|
               &\leq 2\sqrt{3\e}\log(d_A^nd_B^nd_{W_A}) + \eta(2\sqrt{3\e}) \\
               &\leq 2\sqrt{3\e}n\log(d_A^3 d_B) + \eta(2\sqrt{3\e}),
\end{split}\end{equation}
for $\epsilon\leq\frac{1}{12 e^2}$, $\eta(x) = -x \log x$ and
using $d_{W_A}\leq d_A^{2n}$.

Now, using the subadditivity of the von Neumann entropy and the fact
that the overall state is pure we have
\begin{align*}
 H(B^n) + H(C_A) &\geq H(B^nC_A)
                  =    H(W_C\widehat{A}\widehat{B})
                  =    H(W_AR^n)                                          \\
                 &\geq H(W_A) + H(R^n)
                        - 2\sqrt{3\e}n\log(d_A^3 d_B) - \eta(2\sqrt{3\e}) \\
                 &\geq H(A^n) - H(C_A) + H(R^n)
                        - 2\sqrt{3\e}n\log(d_A^3 d_B) - \eta(2\sqrt{3\e}).
\end{align*}
Therefore,
\begin{align*}
   2nQ_A &\geq 2H(C_A)                                           \\
         &\geq H(A^n) - H(B^n) + H(R^n)
                - 2\sqrt{3\e}n\log(d_A^3 d_B) - \eta(2\sqrt{3\e}) \\
         &=    n I(A;R) - 2\sqrt{3\e}n\log(d_A^3 d_B) - \eta(2\sqrt{3\e})
\end{align*}
Dividing by $n$ and letting $\e\rightarrow 0$, we obtain
\begin{equation}
  Q_A \geq \frac{1}{2}I(A;R).
\end{equation}
Switching the roles of Alice and Bob gives the corresponding inequality,
\begin{equation}
  Q_B \geq \frac{1}{2}I(B;R).
\end{equation}
To bound $Q_A+Q_B$ let us return to the situation
where Alive and Bob perform their original encoding.
Then,
\begin{equation}
  H(A^n) = H(C_AW_A) \leq H(W_A) + H(C_A) \leq H(W_A) + nQ_A.
\end{equation}
The first equality follows from the fact that the environment
system is initiated as a pure unentangled state and from the
unitary invariance of the von Neumann entropy.

Combining with the analogous inequality for $B$ leads to,
\begin{equation}
  \label{eqn_4}
  n(Q_A+Q_B) \geq n[H(A)+H(B)]-H(W_A)-H(W_B).
\end{equation}
By similar arguments as before,
\begin{equation}
  | H(W_AW_BR^n) - H(W_AW_B) - H(R^n)|
     \leq 2\sqrt{3\e}n\log(d_A^2d_B^2d_R)+\eta(2\sqrt{3\e}),
\end{equation}
for $\epsilon$ small enough. So,
\begin{equation*}\begin{split}
  H(C_AC_B) &=    H(W_AW_BR^n) \\
            &\geq H(W_A)+H(W_B)-I(W_A;W_B)+H(R^n) \\
            &\phantom{==}
                  - 2\sqrt{3\epsilon}n\log(d_A^2d_B^2d_R)-
\eta(2\sqrt{3\epsilon}).
\end{split}\end{equation*}
Using the purity of the overall state, however, gives
$H(R^n)=nH(AB)$, which combined with the bound $H(C_AC_B)\leq
n(Q_A+Q_B)$, leads to the inequality
\begin{equation}\begin{split}
  \label{eqn_5}
  H(W_A) + H(W_B) &\leq n(Q_A+Q_B) - nH(AB) + I(W_A;W_B) \\
                  &      + 2\sqrt{3\epsilon}n\log(d_A^2d_B^2d_R) +
\eta(2\sqrt{3\epsilon})
\end{split}\end{equation}
Adding equations (\ref{eqn_4}) and (\ref{eqn_5}),
\begin{equation}\begin{split}
  2n(Q_A+Q_B) &\geq n\bigl( H(A)+H(B)+H(AB) \bigr) - I(W_A;W_B) \\
              &\phantom{==}
                   -n\sqrt{3}\epsilon\log(d_A^2d_B^2d_R)-\eta(\sqrt{3}\epsilon).
\end{split}\end{equation}
Thus,
\begin{equation}
  Q_A+Q_B \geq \frac{1}{2}J(A;B) - \frac{1}{n}I(W_A;W_B)
                - 2\sqrt{3\epsilon}\log(d_A^2d_B^2d_R)
                - \frac{\eta(2\sqrt{3\epsilon})}{n}
\end{equation}
Now, let $T:R^n\to R^\prime$ be any CPTP map on
$R^n$. Then we can bound the mutual information $I(W_A;W_B)$
as follows.
\begin{equation*}\begin{split}
  I(W_A;W_B)-I(W_A;W_B|R^\prime)
             &=     \bigl( H(W_A)-H(W_AR^\prime) \bigr)
                   +\bigl( H(W_B)-H(W_BR^\prime) \bigr)               \\
             &\phantom{==}
                   -\bigl( H(W_AW_B)-H(W_AW_BR^\prime) \bigr)
                   -H(R')                                             \\
             &\leq 8\sqrt{3\e}\bigl( \log d_{W_A}+\log d_{W_B}+\log
d_{W_A}d_{W_B} \bigr)
                   + 6 H_2(2\sqrt{3\e})                               \\
             &\leq 8\sqrt{3\e}n \log(d_A^4d_B^4) + 6 H_2(2\sqrt{3\e}),
\end{split}\end{equation*}
where we have used that $W_AW_B$ is almost uncorrelated with $R^\prime$
(via the contractivity of the trace distance under CPTP maps):
\[
  \| \xi^{W_AW_BR^\prime} - \xi^{W_AW_B}\ox\xi^{R^\prime} \|_1 \leq
2\sqrt{3\e},
\]
followed by the Alicki-Fannes inequality~\cite{AF04}. The function
$H_2(x)$ is the binary entropy $H_2(x) = -x \log x - (1-x)\log
(1-x)$. Note that in this way the dimension of $R'$ doesn't enter,
which is desirable as we do not wish to constrain it in any way.

In particular, for small $\e$,
\begin{equation}\begin{split}
  I(W_A;W_B) &\leq I(W_A;W_B|R^\prime)
                    + 8\sqrt{3\e}n \log(d_A^4d_B^4) + 6 H_2(2\sqrt{3\e}) \\
             &\leq I(A^n;B^n|R^\prime)
                    + 8\sqrt{3\e}n \log(d_A^4d_B^4) + 6 H_2(2\sqrt{3\e})
\end{split}\end{equation}
where in the second line we have invoked
the monotonicity of mutual information under local
operations. Therefore,
\begin{equation*}\begin{split}
  Q_A+Q_B &\geq \frac{1}{2}J(A;B) - \frac{1}{2n}I(A^n;B^n|R') \\
          &\phantom{==}
                 - 2\sqrt{3\epsilon}\log(d_A^2d_B^2d_R)
                 - \frac{\eta(2\sqrt{3\epsilon})}{n}
                 - 8\sqrt{3\e}\log(d_A^4d_B^4) - \frac{6H_2(2\sqrt{3\e})}{n}
\end{split}\end{equation*}
By optimising over the CPTP map $T$, we thus obtain
\[\begin{split}
  Q_A+Q_B &\geq \frac{1}{2}J(A;B) - \frac{1}{n} E_{\rm
sq}\bigl((\varphi^{AB})^{\ox n}\bigr)
                 - 10\sqrt{3\e}\log(d_A^4d_B^4) - \frac{7H_2(2\sqrt{3\e})}{n}
\\
          &=    \frac{1}{2}J(A;B) - E_{\rm sq}(\varphi^{AB})
                 - 10\sqrt{3\e}\log(d_A^4d_B^4) - \frac{7H_2(2\sqrt{3\e})}{n},
\end{split}\]
where $E_{\rm sq}(\ph^{AB})$ is the \emph{squashed entanglement} of
$\ph^{AB}$, defined as the infimum of $\smfrac{1}{2} I(A;B|E)$
over extensions $\ph^{ABE}$ of $\ph^{AB}$~\cite{CW04}. We have
used explicitly the fact, proved in the cited paper, that
$E_{\rm sq}(\ph^{\ox n}) = n E_{\rm sq}(\ph)$.

Since $\e>0$ was arbitrary, we have therefore proved the following
outer bound on the achievable rate region:
\begin{theorem}
  \label{thm:distrib-outer}
  The rate region $\region(\ph)$ of fully quantum
  distributed compression of the source $\ph$ is contained in
  the set defined by the inequalities
  \begin{align}
    \label{eqn:outer3}
    Q_A       &\geq \frac{1}{2} I(A;R)_\ph \nonumber \\
    Q_B       &\geq \frac{1}{2} I(B;R)_\ph \\
    Q_A + Q_B &\geq \frac{1}{2} J(A;B)_\ph - E_{\rm sq}(\ph^{AB}). \qquad \qed
\nonumber
  \end{align}
\end{theorem}

In the special case where $\ph^{AB}$ is separable, $E_{\rm sq}(\ph) = 0$,
which implies that the region defined by
Eq.~(\ref{eqn:region}) is optimal. Under certain further technical
assumptions, namely that $\ph^{AB}$ be the density operator of an
ensemble of product pure states satisfying a condition called
irreducibility, the same conclusion was found in~\cite{ADHW04}.
That paper, however, was unable to show that
the bound was achievable.

The appearance of the squashed entanglement in (\ref{eqn:outer3})
may seem somewhat mysterious, but a slight modification of the
protocols based on fully quantum Slepian-Wolf will lead to an
inner bound on the achievable region that is of a similar form.
Specifically, let $D_0(\ph^{AB})$ be the amount of pure state
entanglement that Alice and Bob can distill from $\ph^{AB}$
without engaging in any communication. Since this pure state
entanglement is decoupled from the reference system $R$, they
could actually perform this distillation process and discard the
resulting entanglement before beginning one of their FQSW-based
compression protocols. While neither $I(A;R)$ nor $I(B;R)$ would
change, each of $H(A)$ and $H(B)$ would decrease by
$D_0(\ph^{AB})$. The corresponding inner bound on the achievable
rate region $\region(\ph)$ would therefore be defined by the
inequalities
\begin{align} \label{eqn:inner_final}
  Q_A       &\geq \frac{1}{2} I(A;R)_\ph \nonumber\\
  Q_B       &\geq \frac{1}{2} I(B;R)_\ph \\
  Q_A + Q_B &\geq \frac{1}{2} J(A;B)_\ph - D_0(\ph^{AB}). \nonumber
\end{align}
The only gap between the inner and outer bounds, therefore, is a
gap between different measures of entanglement.

We close this section by exhibiting a class of example sources for
which we believe that the above inner bound is not tight. It is
based on the observation that to arrive at (\ref{eqn:inner_final})
we considered a case where the structure of $W_C$ was very simple.
While in principle $W_C$ could harbor arbitrary tripartite
entanglement with $W_A$ and $W_B$, the decoding for
(\ref{eqn:inner_final}), which is just the FQSW protocol's
decoding, is simply an isometry separating the entanglement with
$R^n$ from that with one, and only one, of $W_A$ and $W_B$. Hence,
we are motivated to try and construct a source that permits Alice
and Bob to extract and discard some ``waste'', such that later on
Charlie can finish off by discarding exactly the purification of
that waste. The purified source is one of the \emph{twisted
states}~\cite{HHHO05} of the form
\[
  \ket{\ph}^{RA'A''B'B''}
       = \sum_{i=1}^d \sqrt{p_i} \ket{i}^{A'}\ket{i}^{B'}
                                 (U_i^{A''B''}\ox I^R)\ket{\phi_0}^{RA''B''},
\]
arbitrary unitaries $U_i$ on the joint system $A''B''$. (It is
understood that $A=A'A''$ and $B=B'B''$.)

Now let us assume that the reduced states $\tau_i^{A''B''} = U_i
\phi_0^{A''B''} U_i^\dagger$ are mutually orthogonal for
$i=1,\ldots,d$. Furthermore, we restrict to the case of
\emph{non-local} unitaries
$U_i$, i.e.~$U_i$ is not a tensor product of local unitaries. 
We conjecture that $D_0(\ph^{AB}) = 0$ or, more specifically, that
because of the nonlocal ``twist'', Alice and Bob cannot extract
pure states from $\ph^{AB}$ by local operations alone. This would
mean that our inner bound yields an achievable rate sum of
\[
  R_A + R_B = \frac{1}{2} J(A;B)
            = H(A) + \frac{1}{2}I(B;R).
\]
However, a better rate sum is attainable because neither Alice nor
Bob need to send the $A'$, $B'$ registers, respectively: if $A''$
and $B''$ are transmitted faithfully, Charlie can coherently
measure $i$, use it to undo $U_i$, so that he is left with the
state $\phi_0^{CR}$. He then has $\ket{i}$ in his waste register
$W_C$, entangled only with the contents of Alice's and Bob's waste
registers $W_A=A'$ and $W_B=B'$. He finishes off by discarding the
waste register, creating $\sum_i \sqrt{p_i}\ket{i}\ket{i}$ afresh
and using a controlled unitary to put back the twist $U_i$ onto
$\phi_0$. Instead of the rates
\[
  R_A = H(A) = H(A'') + H(A'|A''), \quad
  R_B = \frac{1}{2}I(B;R),
\]
they now use strictly less qubit resources,
\[
  R_A' = H(A'') < H(A), \quad
  R_B' = \frac{1}{2}I(B'';R) \leq \frac{1}{2}I(B;R).
\]

\section{On encoding complexity}
\label{sec:efficiency}

While the protocols described so far make use of a unitary
transformation drawn at random according to the Haar measure, that
is not essential. In fact, the only place the Haar measure was
used was in the proof of Lemma \ref{lem:Schur}. Therefore, the
full unitary group could be replaced by any subset yielding the
same average as in the lemma. (We thank Debbie Leung for alerting us
to this possibility.) In fact, DiVincenzo, Leung and Terhal have
shown that
\begin{equation}
\int_{\UU(\CC^{2^n})} (U \ox U) X (U^\dg \ox U^\dg) \, dU
 = \frac{1}{|G_n|} \sum_{g \in G_n} (g \ox g) X (g^\dg \ox g^\dg),
\end{equation}
where $G_n$ is the Clifford group on $n$ qubits~\cite{DLT02}. They
also demonstrate in that paper that choosing an element of $G_n$
from the uniform distribution can be done in time polynomial in
$n$. More specifically, they show that a random walk on a
particular set of generators for $G_n$ mixes in $O(n^8)$ time,
leading to an associated quantum circuit for the selected element
that is of size $O(n^2)$ gates.

Since the Schumacher compression portion of the fully quantum
Slepian-Wolf protocol can also be done in polynomial
time~\cite{CD96}, we conclude that the encoding portion of the
mother can be done efficiently. Since her immediate children,
including entanglement distillation and state merging, are built
by composing the mother with efficient protocols, namely
superdense coding and teleportation, their encodings can also be
found and implemented efficiently.

The transformation from FQSW to the father, however, included
another non-constructive step, namely the choice of a good type
class. Since the number of type classes is polynomial in the
number of qubits in the input, however, that step could also be
implemented efficiently. The corresponding isometries mapping the
shared maximally entangled state and the input space into $A_t$
can also be performed efficiently~\cite{CD96}. Finally, while the proof
presented here implies that the transpose of a random Clifford
group element can be used as the encoding operation, there is in
fact no need for the transpose because the Clifford group is
closed under transposition. Thus, the encoding for the father can
be found and implemented in polynomial time, as can those of his
children, entanglement-assisted classical communication and
quantum communication over a noisy channel.

Finally, because the quantum reverse Shannon protocol consists of
running FQSW backwards in time, there it is Bob's \emph{decoding}
that can be found and implemented efficiently instead of Alice's
encoding.

\section{Discussion} \label{sec:discussion}

We have shown that simple representation-theoretic reasoning,
specifically some quadratic averages, are sufficient to derive the
powerful mother protocol: a fully quantum version of entanglement
distillation with state merging. The mother, in proper mythical
fashion, not only generates her children in the family tree but
also the father protocol and his offspring, the quantum reverse
Shannon theorem, plus an almost complete solution to the
distributed quantum compression problem. We leave it as an open
problem to determine the exact rate region, which we conjecture to
be given by
\begin{align*}
  Q_A     &\geq \frac{1}{2} I(A;R), \quad Q_B \geq \frac{1}{2} I(B;R), \\
  Q_A+Q_B &\geq \frac{1}{2} J(A;B) - F(\ph_{AB}),
\end{align*}
with some functional $F(\ph_{AB})$ of the source density operator.
It is tempting to speculate that $F$, as in our inner and outer bounds
on the rate region, is an entanglement monotone; note that for
separable and for pure states our inner and outer bounds
coincide, giving $0$ and the entropy of entanglement, respectively,
in agreement with the idea that $F$ should be an entanglement measure.

We also note that while we have not pursued the opportunity here,
the one-shot versions of the FQSW, father and reverse Shannon
theorem are natural starting points for developing versions of the
theorem adapted to states or channels with some internal structure
more complicated than i.i.d. It would be interesting to compare
the results of such an effort with the insights of~\cite{BM04} and
\cite{KW05}.

\begin{figure}
 \begin{center}
  \includegraphics[width=5in,bbllx=8,bblly=475,bburx=784,bbury=620]{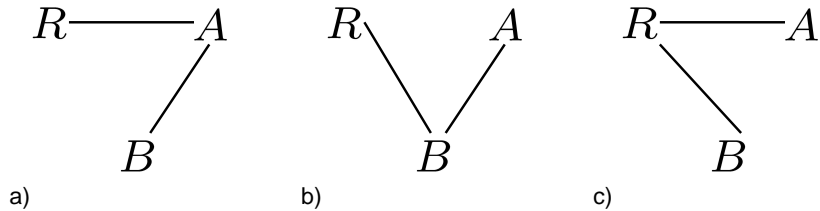}
 \end{center}
 \caption{a) A trivial starting configuration for FQSW. Solid lines represent
 pure state entanglement between two parties. b) The result of Alice sending her Rebecca
 entanglement to Bob. c) The result of Alice sending her Bob entanglement to
 Rebecca.}
 \label{fig:weird}
\end{figure}

We close by highlighting a peculiar feature of the FQSW protocol.
Let $\ket{\ps}$ be a pure state and suppose that Alice-Bob and
Alice-Rebecca both share $n$ copies of $\ket{\ps}$, so that the
global pure state is $\ket{\ph}^{\ox n} = (\ket{\ps}^{A_1
R}\ket{\ps}^{A_2 B})^{\ox n}$. This is a ``trivial'' situation for
FQSW. Instead of using our protocol, Alice can simply transfer her
entanglement with Rebecca to Bob by compressing and sending him
her $A_1$ registers, requiring a rate of $H(A_1) = I(A;R)/2$.
Since Alice and Bob already share $H(A_2) = I(A;B)/2$ ebits of
pure state entanglement, that completes the FQSW protocol. Because
of the symmetry of the situation, the roles of Rebecca and Bob
could also be reversed. Thus, Alice could transfer her Bob
entanglement to Rebecca by Schumacher compressing and sending
$A_2$ to her, requiring a rate $H(A_2) = I(A;B)/2$. It is quite
clear that Alice's system decomposes into an $A_1$ part, which
contains her entanglement with Rebecca, and an $A_2$, which
contains her entanglement with Bob. Note that the entanglement
structure of the final state is very different in the two cases;
see Figure \ref{fig:weird}.
Here's the weirdness: if they use the general FQSW protocol
instead, then since $H(A_1) = H(A_2)$, the \emph{same} unitary
will work in both case with high probability. In other words,
Alice could \emph{first} apply the unitary and \emph{then} decide
whether to transfer her Rebecca entanglement to Bob or her Bob
entanglement to Eve. The only difference in Alice's part of the
protocol is whether she sends the qubits (at rate arbitrarily
small above $H(A_1)$) to Bob or to Rebecca. Thus, the localization
of the entanglement so evident in the trivial implementation of
the protocol disappears in the general implementation. The same
subsystem can be made to carry both forms of entanglement
simultaneously, compatible with either recipient!

\subsection*{Acknowledgments}
The authors would very much like to thank Debbie Leung for
bringing to their attention the possibility of replacing Haar
measure unitaries with random Clifford group elements. They would
also like to thank Isaac Chuang, Ignacio Cirac, Fr\'ed\'eric
Dupuis, Renato Renner and J\"urg Wullschleger for their helpful
comments. AA appreciates the support of the US National Science
Foundation through grant no. EIA-0086038. ID was partially
supported by the NSF under grant no. CCF-0524811. PH was supported
by the Canada Research Chairs program, the Canadian Institute for
Advanced Research, and Canada's NSERC. He is also grateful to the
Benasque Centre for Science and CQC Cambridge for their
hospitality. AW was supported by the U.K. Engineering and Physical
Sciences Research Council's ``IRC QIP'', and by the EC projects
RESQ (contract IST-2001-37759) and QAP (contract IST-2005-15848),
as well as by a University of Bristol Research Fellowship.

\appendix
\section{Properties of typical and type projectors.}
We present here a number of consequences of the method of type classes.
 Denote by $x^n$ a sequence $x_1x_2\dots x_n$, where each
$x_i$ belongs to the finite set $\cX$. Denote by $|\cX|$ the
cardinality of $\cX$. Denote by $N(x|x^n)$ the number of
occurrences of the symbol $x$ in the sequence $x^n$. The
\emph{type} $t^{x^n}$ of a sequence $x^n$ is a probability vector
with elements $t^{x^n}_x = \frac{N(x_i|x^n)}{n}$.
Denote the set of sequences of type $t$ by
$$\cT^n_t=\{x^n\in \cX^n: t^{x^n}=t\}.$$
For the probability distribution $p$ on the set $\cX$ and $\delta >0$,
let $\tau_\delta=\{t:\forall x \in \cX,\ |t_x-p_x|\leq \delta\}$. 
$|\tau_\delta|=a$.
Define the set of $\delta$-typical sequences of length $n$ as 
$\cT^{n}_{p,\delta}$, as
\beq
\cT^{n}_{p,\delta} =\underset{t \in \tau_\delta}{\bigcup} \cT^n_t
                   =\{x^n:\forall x\in\cX,\ |t^{x^n}_x-p_x|\leq \delta\}.
\eeq
Define the probability distribution $p^n$ on $\cX^n$ to be the
$n$-fold product of $p$. The sequence $x^n$ is drawn from $p^n$ if
and only if each letter $x_i$ is drawn independently from $p$.
Typical sequences enjoy many useful properties \cite{CT91,CK81}.
Let $H(p)
= - \sum_x p_x \log p_x$ be the Shannon entropy of $p$. For any
$\epsilon,\delta>0$, and all sufficiently large $n$ for which \beq
p^n(\cT_{p,\delta}^{n})  \geq  1-\epsilon   \label{cc1} \eeq \beq
2^{-n[ H(p)+c\delta ]}  \leq  p^n(x^n)  \leq   2^{-n[
H(p)-c\delta]},  \,\,\, \forall x^n \in \cT_{p,\delta}^{n}
\label{cc2} \eeq \beq (1 - \epsilon)^{-1}  2^{n[ H(p)-c\delta]}
\leq  |\cT_{p,\delta}^{n}| \leq 2^{n[ H(p)+c\delta]}, \label{cc3}
\eeq for some constant $c$. For $t\in \tau_\delta$ and for
sufficiently large $n$, the cardinality $D_t$ of $\cT^n_t$ is
bounded as \cite{CT91}
\beq \label{cc4} D_t \geq 2^{n [H(p) - \iota(\delta)]} \eeq and
the function $\iota(\delta) \rightarrow 0$ as $\delta \rightarrow
0$.

The above concepts generalize to the quantum setting by virtue of the spectral
theorem. Let $\rho=\sum_{x\in\cX}p_x \ket{x}\bra{x}$ be the
spectral decomposition of a given  density matrix $\rho$.
In other words, $\ket{x}$ is the eigenstate of
$\rho$ corresponding to eigenvalue $p_x$.
The von Neumann entropy of the density matrix $\rho$ is
$$H(\rho)=-\Tr{\rho\log\rho}=H(p).$$
Define the type projector $$\Pi^n_t=\sum_{x^n\in\cT^n_{t}}\ket{x^n}\bra{x^n}.$$
The typical subspace associated with the density matrix $\rho$
is defined as
$$
\Pi^n_{\rho, \delta}
=\sum_{x^n\in\cT_{p,\delta}^n}\ket{x^n}\bra{x^n}=\sum_{t\in\tau_{\delta}}\Pi^n_t
.
$$
Properties analogous to (\ref{cc1}) -- (\ref{cc4}) hold.
For any $\epsilon,\delta>0$, and all sufficiently large $n$ for which
\beq
\Tr{\rho^{\otimes n} \Pi^n_{\rho, \delta}}  \geq  1-\epsilon  \label{q1}
\eeq
\beq
2^{-n[ H(\rho)+c\delta]}\Pi^n_{\rho, \delta}  \leq  \Pi^n_{\rho, \delta}
\rho^{\otimes n} \Pi^n_{\rho, \delta}
  \leq 2^{-n[ H(\rho)-c\delta ]}\Pi^n_{\rho, \delta}
\label{q2}, \eeq \beq (1 - \epsilon)^{-1}  2^{n[ H(\rho)-c\delta]}
\leq  \Tr{\Pi^n_{\rho, \delta}} \leq 2^{n[ H(\rho)+c\delta]},
\label{q3} \eeq for some constant $c$. For $t\in \tau_\delta$ and
for sufficiently large $n$, the support dimension of the type
projector $\Pi^n_t$ is bounded as \beq \label{cc7} \Tr\Pi^n_t \geq
2^{n [H(\rho) - \iota(\delta)]}. \eeq

Henceforth we shall drop the $n$ and $\delta$ indices. In dealing
with a multiparty system such as $\ket{\ps'} =
(\ket{\ph}^{ABR})^{\ox n}$, we shall label the typical projectors
corresponding to the various sybsystems by $\Pi_A$ etc. A variant
of the gentle measurement lemma \cite{W99} states that if $\Tr \Pi
\rho \geq 1 - \epsilon$ then $\| \rho - \widehat{\sigma} \|_1 \leq
2 \sqrt{\epsilon}$, where $\widehat{\sigma} = \sigma /
\Tr{\sigma}$ and $\sigma = \Pi \rho \Pi$. Applying it together
with  (\ref{q1}) gives
$$
\| \psi' - \Pi_A \psi' \Pi_A / (\Tr \psi' \Pi_A )\|_1 \leq 2 \sqrt{\epsilon}.
$$
The Schumacher  compression operation $\cE$ projects onto
$\Pi_A$ with probability $\Tr {\psi'} \Pi_A \geq 1 - \epsilon$.
Thus
$$
\|\cE (\psi') - \Pi_A \psi' \Pi_A/ (\Tr \psi' \Pi_A ) \|_1 \leq 2 \epsilon.
$$
The triangle inequality now gives
$$
\|\cE (\psi') -  \psi'  \|_1 \leq 2 \epsilon + 2 \sqrt{\epsilon}.
$$
Define $\ket{\psi}$ to be the normalized version of
the state
\begin{equation}
(\Pi_A \ox \Pi_B \ox \Pi_R) \ket{\ps'}.
\end{equation}
Since $\Pi_R$,  $\Pi_A$ and  $\Pi_B$ commute,
they satisfy a sort of union bound,
\begin{equation}
\Pi_A \otimes \Pi_B \otimes \Pi_R \geq
 \Pi_A  + \Pi_B  + \Pi_R - 2 I.
\label{union}
\end{equation}
Combining this with the same variant of the gentle measurement
lemma as before and (\ref{q1}) gives
$$
\| \psi' - \psi \|_1 \leq 2 \sqrt{3 \epsilon}.
$$
Observe
$$
\Pi_A {\psi'}^{ABR} \Pi_A
\geq \Pi_A {\psi'}^{ABR}  (\Pi_B \ox \Pi_R) {\psi'}^{ABR} \Pi_A.
$$
Then
\begin{equation}
\begin{split}
\Pi_A {\psi'}^{A} \Pi_A &=
\Tr_{BR}[\Pi_A {\psi'}^{ABR} \Pi_A] \\
& \geq \Tr_{BR}[\Pi_A {\psi'}^{ABR}  (\Pi_B \ox \Pi_R) {\psi'}^{ABR} \Pi_A] \\
&=  \Tr_{BR}[(\Pi_A \ox \Pi_B \ox \Pi_R) {\psi'}^{ABR} (\Pi_A \ox \Pi_B \ox
\Pi_R)] \\
& \geq (1 - 3 \epsilon) \psi^A.
\end{split}
\end{equation}
Combining with inequalities (\ref{q2}) and (\ref{q3})
gives
$$
\Tr[(\psi^A)^2] \leq (1 - 3 \epsilon)^{-1} 2^{-n [H(A) - c \delta]}.
$$
Define $P'_t = \Tr \psi' \Pi^t_A$ and $P_t = \Tr \psi \Pi^t_A$. By
(\ref{q2}) and (\ref{cc7}), $P'_t  \geq  2^{- n[c \delta +
\iota(\delta)]}$ for all $t \in \tau_\delta$. Define
$\ket{\psi_t'}$ and $\ket{\psi_t}$ to be the
 normalized  versions of the states  ${\Pi^t_A \ket{\psi'}}$
and $\Pi^t_A \ket{\psi}$, respectively.
Since $\Pi_A \ket{\psi'} = \sum_{t\in \tau_\delta} \sqrt{P_t'}  \ket{\psi_t'}$,
and $\ket{\psi} = \sum_{t\in \tau_\delta}
\sqrt{P_t}  \ket{\psi_t}$, we have
\begin{equation}
\sum_{t\in \tau_\delta} \sqrt{P_t P'_t} |\bra{\psi_t}\psi_t'\rangle|
 \geq
 |\bra{\psi}\psi'\rangle| \geq {1 - 3 \epsilon}.
\end{equation}
We now claim that there exists a $t$ for which both
\begin{equation}
|\bra{\psi_t}\psi_t'\rangle| \geq {1 - 18 \epsilon}
 \label{tree}
\end{equation}
and $P_t \geq \frac{1}{3} P_t' \geq \frac{1}{3} 2^{- n[c \delta +
\iota(\delta)]}$. First, by Cauchy-Schwarz,
$$
\sum_t \frac{1}{2}(P_t + P_t') |\bra{\psi_t}\psi_t'\rangle| \geq 1 - 3 \epsilon,
$$
so that
$$
\sum_t P_t' |\bra{\psi_t}\psi_t'\rangle| \geq 1 - 6 \epsilon.
$$
Thinking of $P'_t$ as a probability distribution over $t$, the
probability that $P_t' > {3} P_t$ is upper bounded by
$\frac{1}{3}$, as is the probability that
$|\bra{\psi_t}\psi_t'\rangle| \leq 1 - 18 \epsilon$. Hence, there
exists a $t$ for which both events are false, yielding the claim.
Choose $t$ to be one that satisfies the claim. Then
$$
\| \psi_t - \psi_t' \|_1 \leq 12 \sqrt{\epsilon}.
$$
From
$$
\Tr_{AR} [(\Pi_A^t \ox \Pi_B \ox \Pi_R) {\psi'}^{ABR} (\Pi_A^t \ox \Pi_B \ox
\Pi_R)]
\leq \Tr_{AR} [(\Pi_A \ox \Pi_B \ox \Pi_R) {\psi'}^{ABR} (\Pi_A \ox \Pi_B \ox
\Pi_R)]
$$
and  $\Tr\Pi_A^t\psi  \geq \frac{1}{3} 2^{- n[c \delta +
\iota(\delta)]}$ it follows that
$$
\Tr[(\psi_t^B)^2] \leq {3} \cdot 2^{ n[c \delta + \iota(\delta)]}
\Tr[(\psi^B)^2] \leq  3 \cdot (1 - 3 \epsilon)^{-1} 2^{-n [H(B) -
2c \delta - \iota(\delta)]}.
$$
A similar bound holds for $\Tr[(\psi_t^R)^2]$.

Thus we have shown properties i)-iv)  of Section  \ref{sec:iid}
and i)-v)  of Section  \ref{sec:father:iid}.

\bibliographystyle{unsrt}
\bibliography{fqsw}

\begin{thebibliography}{10}

\bibitem{H98}
A.~S. Holevo.
\newblock The capacity of the quantum channel with general signal states.
\newblock {\em IEEE Trans. Inf. Theory}, 44:269--273, 1998.

\bibitem{SW97}
B.~Schumacher and M.~D. Westmoreland.
\newblock Sending classical information via noisy quantum channels.
\newblock {\em Phys. Rev. A}, 56:131--138, 1997.

\bibitem{BSST99}
C.~H. Bennett, P.~W. Shor, J.~A. Smolin, and A.~V. Thapliyal.
\newblock Entanglement-assisted classical capacity of noisy quantum channels.
\newblock {\em Phys. Rev. Lett.}, 83:3081, 1999.
\newblock ar{X}iv.org:quant-ph/9904023.

\bibitem{BSST02}
C.~H. Bennett, P.~W. Shor, J.~A. Smolin, and A.~V. Thapliyal.
\newblock Entanglement-assisted capacity of a quantum channel and the reverse
  {S}hannon theorem.
\newblock {\em {IEEE} Trans. Inf. Theory}, 48(10):2637, 2002.
\newblock ar{X}iv.org:quant-ph/0106052.

\bibitem{L96}
S.~Lloyd.
\newblock Capacity of the noisy quantum channel.
\newblock {\em Phys. Rev. A}, 55:1613, 1996.

\bibitem{S02}
P.~W. Shor.
\newblock The quantum channel capacity and coherent information.
\newblock Lecture notes, {MSRI} workshop on quantum computation, 2002.
\newblock Available online at
  http://www.msri.org/publications/ln/msri/2002/quantumcrypto/shor/1/.

\bibitem{D05}
I.~Devetak.
\newblock The private classical capacity and quantum capacity of a quantum
  channel.
\newblock {\em {IEEE} Trans. Inf. Theory}, 51(1):44, 2005.
\newblock ar{X}iv.org:quant-ph/0304127.

\bibitem{DW05}
I.~Devetak and A.~Winter.
\newblock Distillation of secret key and entanglement from quantum states.
\newblock {\em Proc. R. Soc. Lond. A}, 461:207--237, 2005.
\newblock ar{X}iv.org:quant-ph/0306078.

\bibitem{HHHLT01}
M.~Horodecki, P.~Horodecki, R.~Horodecki, D.~W. Leung, and B.~M. Terhal.
\newblock Classical capacity of a noiseless quantum channel assisted by noisy
  entanglement.
\newblock {\em Quant. Inf. Comp.}, 1:70--78, 2001.
\newblock ar{X}iv.org:quant-ph/0106080.

\bibitem{H04}
A.~W. Harrow.
\newblock Coherent communication of classical messages.
\newblock {\em Phys. Rev. Lett.}, 92:097902, 2004.
\newblock ar{X}iv.org:quant-ph/0307091.

\bibitem{DHW04}
I.~Devetak, A.~W. Harrow, and A.~Winter.
\newblock A family of quantum protocols.
\newblock {\em Phys. Rev. Lett.}, 93:230504, 2004.
\newblock ar{X}iv.org:quant-ph/0308044.

\bibitem{HOW05}
M.~Horodecki, J.~Oppenheim, and A.~Winter.
\newblock Partial quantum information.
\newblock {\em Nature}, 436:673--676, 2005.
\newblock ar{X}iv.org:quant-ph/0505062.

\bibitem{HOW05b}
M.~Horodecki, J.~Oppenheim, and A.~Winter.
\newblock Quantum state merging and negative information.
\newblock ar{X}iv.org:quant-ph/0512247, 2005.

\bibitem{D05b}
I.~Devetak.
\newblock A triangle of dualities: reversibly decomposable quantum channels,
  source-channel duality, and time reversal.
\newblock ar{X}iv.org:quant-ph/0505138, 2005.

\bibitem{DHW05}
I.~Devetak, A.~W. Harrow, and A.~Winter.
\newblock A resource framework for quantum {S}hannon theory.
\newblock ar{X}iv.org:quant-ph/0512015, 2005.

\bibitem{SW71}
D.~Slepian and J.~K. Wolf.
\newblock Noiseless coding of correlated information sources.
\newblock {\em {IEEE} Trans. Inf. Theory}, 19:461--480, 1971.

\bibitem{YHD06}
J.~Yard, P.~Hayden, and I.~Devetak.
\newblock Quantum broadcast channels.
\newblock ar{X}iv.org:quant-ph/0603098, 2006.

\bibitem{YDH05}
J.~Yard, I.~Devetak, and P.~Hayden.
\newblock Capacity theorems for quantum multiple access channels-part {I}:
  {C}lassical-quantum and quantum-quantum capacity regions.
\newblock ar{X}iv.org:quant-ph/0501045, 2005.

\bibitem{SVW05}
J.~A. Smolin, F.~Verstraete, and A.~Winter.
\newblock Entanglement of assistance and multipartite state distillation.
\newblock {\em Phys. Rev. A}, 72(5):052317, 2005.
\newblock ar{X}iv.org:quant-ph/0505038.

\bibitem{GPW05}
B.~Groisman, S.~Popescu, and A.~Winter.
\newblock On the quantum, classical and total amount of correlations in a
  quantum state.
\newblock {\em Phys. Rev. A}, 72:032317, 2005.
\newblock ar{X}iv.org:quant-ph/0410091.

\bibitem{U76}
A.~Uhlmann.
\newblock The `transition probability' in the state space of a {$^*$}-algebra.
\newblock {\em Rep. Math. Phys.}, 9:273, 1976.

\bibitem{BDHSW06}
C.~H. Bennett, I.~Devetak, A.~W. Harrow, P.~W. Shor, and A.~Winter.
\newblock The {Q}uantum {R}everse {S}hannon {T}heorem.
\newblock In preparation, 2006.

\bibitem{FvG99}
C.~A. Fuchs and J.~van~de Graaf.
\newblock Cryptographic distinguishability measures for quantum mechanical
  states.
\newblock {\em IEEE Trans. Inf. Theory}, 45:1216--1227, 1999.

\bibitem{F73}
M.~Fannes.
\newblock A continuity property of the entropy density for spin lattice
  systems.
\newblock {\em Commun. Math. Phys.}, 31:291--294, 1973.

\bibitem{AF04}
R.~Alicki and M.~Fannes.
\newblock Continuity of quantum conditional information.
\newblock {\em J. Phys. A}, 37:L55--L57, 2004.
\newblock ar{X}iv.org:quant-ph/0312081.

\bibitem{CW04}
M.~Christandl and A.~Winter.
\newblock Squashed entanglement - {A}n additive entanglement measure.
\newblock {\em J. Math. Phys.}, 45(3):829--840, 2004.
\newblock ar{X}iv.org:quant-ph/0308088.

\bibitem{ADHW04}
C.~Ahn, A.~Doherty, P.~Hayden, and A.~Winter.
\newblock On the distributed compression of quantum information.
\newblock {IEEE} {T}rans. {I}nf. {T}heory, to appear.
  ar{X}iv.org:quant-ph/0403042, 2004.

\bibitem{HHHO05}
K.~Horodecki, M.~Horodecki, P.~Horodecki, and J.~Oppenheim.
\newblock Secure key from bound entanglement.
\newblock {\em Phys. Rev. Lett.}, 94(16):160502, 2005.
\newblock ar{X}iv:quant-ph/0309110.

\bibitem{DLT02}
D.~P. DiVincenzo, D.~W. Leung, and B.~M. Terhal.
\newblock Quantum data hiding.
\newblock {\em {IEEE} Trans. Inf. Theory}, 48(3):580--598, 2002.
\newblock ar{X}iv.org:quant-ph/0103098.

\bibitem{CD96}
R.~Cleve and D.~P. Di{V}incenzo.
\newblock Schumacher's quantum data compression as a quantum computation.
\newblock {\em Phys. Rev. A}, 54(4):2636--2650, 1996.
\newblock ar{X}iv.org:quant-ph/9603009.

\bibitem{BM04}
G.~Bowen and S.~Mancini.
\newblock {Quantum channels with a finite memory}.
\newblock {\em Phys. Rev. A}, 69(1):12306, 2004.
\newblock ar{X}iv.org:quant-ph/0305010.

\bibitem{KW05}
D.~Kretschmann and R.F. Werner.
\newblock {Quantum channels with memory}.
\newblock {\em Phys. Rev. A}, 72(6):62323, 2005.
\newblock ar{X}iv.org:quant-ph/0502106.

\bibitem{CT91}
T.~M. Cover and J.~A. Thomas.
\newblock {\em Elements of Information Theory}.
\newblock Wiley, 1991.

\bibitem{CK81}
I.~Csiszar and J.~K\"orner.
\newblock {\em Information Theory: Coding Theorems for Discrete Memoryless
  Systems}.
\newblock Academic Press, 1981.

\bibitem{W99}
A.~Winter.
\newblock Coding theorem and strong converse for quantum channels.
\newblock {\em IEEE Trans. Inf. Theory}, 45(7):2481--2485, 1999.

\end{thebibliography}

\end{document}